\documentclass[acmnow]{acmtrans2m}

\usepackage{theomac}
\usepackage{graphicx}
\usepackage{graphicx}
\usepackage{caption}
\usepackage{subcaption}

\newtheorem{theorem}{Theorem}[section]
\newtheoremWithMacro{rtheorem}[theorem]{Theorem}

\newtheorem{corollary}[theorem]{Corollary}
\newtheorem{proposition}[theorem]{Proposition}
\newtheoremWithMacro{rproposition}[theorem]{Proposition}

\newdef{definition}[theorem]{Definition}
\newdef{remark}[theorem]{Remark}

\usepackage[utf8]{inputenc}
\usepackage{amsmath}
\usepackage{amssymb}
\usepackage{algorithm}
\usepackage{algpseudocode}
\newtheorem{fact}{Fact}
\newtheorem{example}{Example}

\newcommand{\cGAP}{\emph{cGAP}}
\newcommand{\cGAPs}{\emph{cGAP}s}
\newcommand{\eqs}{{\tt SE}}
\newcommand{\SE}{strong equilibrium}
\newcommand{\SEs}{strong equilibria}
\newcommand{\G}{{\cal G}}
\newcommand{\atoms}{{\tt atoms}}
\newcommand{\Var}{{\cal V}}
\newcommand{\AVar}{{\tt AVar}}
\newcommand{\la}{{l}}
\newcommand{\V}{{\tt V}}
\newcommand{\E}{{\tt E}}

\def\choicegaps{{\textsf{ChoiceGAP}s}}
\def\choicegap{{\textsf{ChoiceGAP}}}
\def\facebook{{\textsf{FaceBook}}}
\def\flickr{{\textsf{Flickr}}}

\def\coh{{\textsf{coh}}}
\def\calp{{\mathcal P}}
\def\calpv{{\mathcal P}_v}

\title{ChoiceGAPs: Competitive Diffusion as a Massive Multi-Player Game in Social Networks}
\author{Edoardo Serra, Francesca Spezzano and V.S. Subrahmanian}

\begin{abstract}
We consider the problem of modeling competitive diffusion in real world social networks via the notion of \choicegaps\ which combine choice logic programs due to Sacc{\`a} and Zaniolo and Generalized Annotated Programs due to Kifer and Subrahmanian.  We assume that each vertex in a social network is a player in a multi-player game (with a huge number of players) --- the choice part of the \choicegaps\ describe utilities of players for acting in various ways based on utilities of their neighbors in those and other situations. We define multi-player Nash equilibrium for such programs --- but because they require some conditions that are hard to satisfy in the real world, we introduce a new  model-theoretic concept of strong equilibrium.  We show that stable equilibria can capture all Nash equilibria. 
We prove a host of complexity (intractability) results for checking existence of strong equilibria (as well as related counting complexity results), together with algorithms to find them. We then identify a class of \choicegaps\ for which stable equilibria can be polynomially computed. We develop algorithms for computing these equilibria under various restrictions. We come up with the important concept of an estimation query which can compute quantities w.r.t. a given strong equilibrium, and approximate ranges of values (answers) across the space of strong equilibria. Even though we show that computing range answers to estimation queries exactly is intractable, we are able to identify classes of estimation queries that can be answered in polynomial time. We report on experiments we conducted with a real-world \facebook\ data set surrounding the 2013 Italian election showing that our algorithms have good predictive accuracy with an Area Under a ROC Curve that, on average, is over 0.76.
\end{abstract}




\begin{document}

\begin{bottomstuff}
Edoardo Serra, Francesca Spezzano and V.S. Subrahmanian are with the University of Maryland Institute for Advanced Computer Studies (UMIACS), College Park, MD 20742, USA. 
\end{bottomstuff}

  \maketitle

\section{Introduction}
The need to understand and predict the results of diffusion in social networks has taken on great importance in recent years. Most past work assumes a non-competitive scenario in which we model diffusion of one phenomenon at a time. However, in the real world, multiple competing phenomena are often diffusing concurrently. For instance, the ``likes'' for a political candidate $A$ and a competing candidate $B$ might be mutually exclusive --- a person may support at most one of them. Likewise, two competing marketing campaigns, one each for the iOS and Android platforms, may garner ``likes'' from supporters of each but it is unlikely that they will both get ``likes'' from the same person. In a similar vein, various ``issues'' may have supporters - for instance, in the US, there are ``pro'' and ``anti'' abortion supporters, ``pro'' and ``anti'' immigration supporters, ``pro'' and ``anti'' Obamacare (Mr. Obama's 2014 health care law) supporters, and so forth. In all these cases, people typically choose at most one of these \emph{positions.}

In this paper, we present the \choicegap\ framework using which we can model competing diffusive processes via a mix of generalized annotated programs (GAPs) due to Kifer and Subrahmanian~\cite{ks92} and Choice Logic Programs proposed by Sacc{\`a} and Zaniolo~\cite{SacZan90}. The use of GAPs to model diffusion processes was already proposed in 
\cite{SNDOP_query,BroechelerSS10} --- there, the authors show how many well-known diffusion models can be expressed as GAPs.\footnote{Specifically, the authors  of \cite{SNDOP_query} show that they can express cascade models such as \cite{ChaMisloveGummadi_2009} used to model the spread of ``favorites'' in \flickr, tipping models such as the Jackson-Yariv model of product adoption in economics \cite{jy05}, the SIR and the SIS models of disease spread~\cite{anderson79,Hethcote76}, as well as homophilic models such as those involving mobile phone usage~\cite{aral09}. Thus, GAPs can be used to represent many different diffusion models and hence we do not repeat these representations here.}
\cite{SNDOP_query} assumes only one diffusive process is occurring at a time and that there is no competition going on. \cite{BroechelerSS10} presents first steps toward modeling competitive diffusion but do so by identifying one solution of a convex set of nonlinear (conic) constraints. They do not present complexity results, nor do they present accuracy results based on real data. Moreover, their framework usually
takes hours to compute.

When members of a social network can choose at most one of $n$ different competing positions, we first use $n$ different diffusion models, each capturing how support for each of the competing positions spreads through a social network. Each vertex in the social network can be considered to be a player in the game with one of $n+1$ choices to make. He can either choose one of the competing positions or he can choose none -- he can never choose more than one. The players' utility (for a given course of action, e.g. being pro-Obamacare) is defined by a choice rule that uses inputs from the $n$ different GAPs. We can think of the spread of support for a political candidate in a social network as a ``game'' in which the Nash equilibrium~\cite{nash50} represents a stable adoption of positions by the members of the social network. When a Nash equilibrium occurs, no player has an incentive to deviate from his current position.   

After introducing the syntax and semantics of \choicegaps\ in Section~\ref{sec:choicegaps}, we
formally define our game and the notions of ``strong'' and ``Nash'' equlibria in Section~\ref{sec:game-formalization}. Unfortunately, Nash equilibria require assumptions that are unrealistic in real world social networks. We show that every game (in our sense) can be expressed using a \choicegap\ in such a way that the Nash equilibrium and the strong equilibrium coincide --- thus, strong equilibria capture all the good points of Nash equilibria (for our games) without inheriting any of the disadvantages.
We then study the complexity of various problems relating to these equilibria in Section~\ref{sec:complexity}. Specifically, and perhaps unsurprisingly, we show that these problems are computationally intractable. 

Then, in Section~\ref{sec:ilp-equilibria}, we provide mixed integer linear programs (MILPs for short) to capture coherent models of \choicegaps\ and then use these MILPs in an algorithm to compute all strong equilibria. 
Because Section~\ref{sec:complexity} shows that problems relating to the concept of our equilibra are intractable, we identify a class of \choicegaps\ called
``Vertex Independent Choice'' (or VIC) programs and show that for a class of these called $VIC_2$ programs, we can find a strong equilibrium in polynomial time (and they are guaranteed to exist). However, the problem of
finding all such equilibria is not likely to be polynomially computable. Section~\ref{sec:comp-diff-est} looks at ``queries'' that a user may wish to express in such social network games -- typically these queries compute expected diffusions of competing concepts.
We develop algorithms to answer such queries under different assumptions. 

Section~\ref{sec:expts} describes experiments we have carried out pertaining to a competitive situation during the 2013 Italian election using data gathered from \facebook.
We considered 4 different ``competitions'' and tested our algorithms with various combinations of diffusion models. We measured accuracy of our algorithms using the standard ``Area Under ROC Curve'' or AUROC metric. A receiver operating characteristic (ROC) curve plots false positive rates on the $x$-axis and true-positive rates on the $y$-axis. The AUROC is the area under the resulting curve. An AUROC of 0.5 shows the algorithm has no predictive power.  Depending upon the settings used in our algorithms, our experiments show that our algorithms  achieve an avverage AUROC of 0.762, showing good predictive accuracy.

\section{A Simple Motivating Example}\label{sec:example}
Suppose we want to study an election (say in the UK) between Labour
and Tories using a social network such as \facebook.
Adoption of ``likes'' on \facebook\ may diffuse through the network in accordance with a diffusion model for each party as shown below.\footnote{We will give a formal definition of diffusion models later.}

$$\mathsf{DM\_Labour} = \left\{
\begin{array}{lcl}
voteLabour(A):X & \leftarrow & suptBrown(A):X\\
voteLabour(B):X & \leftarrow & voteLabour(A):X, \ knows(B,A):1\\
voteLabour(B):X & \leftarrow & voteLabour(A):X, \ mentor(B,A):1,\\
& & student(B):1\\
voteLabour(B):X & \leftarrow & voteLabour(A):X, \ olderRel(B,A):1
\end{array}\right.
$$

$$\mathsf{DM\_Tory} = \left\{
\begin{array}{lcl}
voteTory(A):X & \leftarrow & likeCam(A):X\\
voteTory(B):X & \leftarrow & voteTory(A):X, \ mentor(B,A):1,\\
& & employee(B):1\\
voteTory(B):X & \leftarrow & voteTory(A):X, \ idol(B,A):1,\\
& & young(B):1
\end{array}\right.
$$
Here, $voteLabour(A):X$ and $voteTory(A):Y$ denote the probabilities ($X,Y$) of the vertex $A$
voting for Labour, resp. Tory,
with value $X$ and $Y$, respectively. The second rule in $\mathsf{DM\_Labour}$ can intuitively be read as
``If $B$ knows $A$ with 100\% certainty and $A$ will vote for Labour with $X$\% certainty, then $B$ will vote Labour with $X$\% certainty.

\emph{Conflicting predicate symbols are specified explicitly in advance by the user or application.} 
For instance, $voteLabour()$ and $voteTory()$ are conflicting predicate symbols because a user may have at most one of these two properties.\footnote{Throughout this paper, we assume
that conflicting predicate symbols are always unary, denoting a property of a user (vertex), e.g. vote-Labour vs. voteTory or pro-Obamacare vs. anti-Obamacare.}
 In contrast, predicates like $mentor()$ and $idol()$ may not be conflicting.
Consequently, each person (vertex in the network) must choose the party to vote for according to his maximal
utility, and then spread ``good'' vibes about his chosen party through the network. 

For each conflicting vertex predicate $vp$, we introduce a
"utility" predicate $vp^U$ and a "decision" predicate $vp^D$. The former
contains the utility value of the corresponding choice, while
the latter represents the vertex's actual choice. Given a set of
conflicting predicate symbols, for each vertex, only one corresponding decision atom
must have a value greater than $0$ and equal to the value of the corresponding
utility atom, while others must have value $0$. A \choicegap\ (to be formally defined later) characterizing this situation is shown below. Such \choicegaps\ can be automatically constructed once the set of
conflicting predicates is specified and once the diffusion models for those individual predicates is known.

$$
\begin{array}{rcl}
voteLabour^U(A):X & \leftarrow & suptBrown(A):X\\
voteLabour^U(B):X & \leftarrow & voteLabour^{D}(A):X, \ knows(B,A):1\\
voteLabour^U(B):X & \leftarrow & voteLabour^{D}(A):X, \ mentor(B,A):1,\\
& & student(B):1\\
voteLabour^U(B):X & \leftarrow & voteLabour^{D}(A):X, \ olderRel(B,A):1\\
voteTory^U(A):X & \leftarrow & likeCam(A):X\\
voteTory^U(B):X & \leftarrow & voteTory^{D}(A):X, \ mentor(B,A):1,\\
& & employee(B):1\\
voteTory^U(B):X & \leftarrow & voteTory^{D}(A):X, \ idol(B,A):1,\\
& & young(B):1\\
voteTory^{D}(X),voteLabour^{D}(X) & \hookleftarrow &
voteTory^U(X),voteLabour^U(X)
\end{array}
$$
Each competing predicate symbol $voteLabour$ and $voteTory$ in our example diffusion models $\mathsf{DM\_Labour, DM\_Tory}$ respectively, thus has the associated predicate symbols
$voteLabour^U$ and $voteLabour^D$.

When constructing a \choicegap\ from a set of diffusion models, the 
conflicting predicates appearing in a rule head are replaced by the
corresponding utility predicates, while
the conflicting predicates appearing in the body are substituted by the
corresponding decision predicates.
Thus, the utility atoms for a vertex directly depend on the decision
atoms of other vertexes. For instance,
in the second rule above, we have $voteLabor^U(B):X\leftarrow voteLabor^D(A):X, knows(B,A):1.$ This rule says that if $A$ decides to vote labor with certainty $X$ and
$B$ knows $A$ for sure (100\% certainty) then the utility to $B$ of voting for labor is also $X$. \emph{Note that when individual diffusion models (which have no utility/decision predicates) are combined, the resulting
\choicegaps\ have such predicates clearly distinguishing utility predicates and decision predicates.}

The last rule of the above \choicegap\  is a ``new type" of rule, called a \emph{Vertex Choice rule} (VC-rule for short), requiring that a vertex can only
 vote for one party. The body of the rule indicates
the possible choices for a vertex, while the head contains the possible decisions.
This rule imposes the fact that each vertex must choose according to utility
values reported in its utility atoms and make a single choice, i.e. only one decision atom for that vertex is true.

Roughly speaking, each vertex operates like a diffusion model filter by
deciding to spread one of the competing diffusion models --- this model is selected according to
the utility function. The vertex does not spread any other competing predicate.

\begin{example}\em
Suppose Tom is a student
($student(Tom):1$) who likes Cameron with certainty $0.6$ ($likeCam(Tom):0.6$),
but who has a mentor John ($mentor(John,$ $Tom):1$) who votes for Labour with certainty $0.8$ ($voteLabour(John):0.8$).
By the rules in the previous \choicegaps\, Tom's utilities are as follows: $voteTory^U(Tom):0.6$
or $voteLabour^U(Tom):0.8$. As Tom's utility is bigger if he votes for Labour,
the right choice for him is to vote for Labour: $voteLabour^D(Tom):0.8$ and
$voteTory^D(Tom):0$.
Thus, Tom diffuses his choice through the Labour diffusion model $\mathsf{DM\_Labour}$
to his neighbors. \hfill$\Box$
\end{example}

\section{Choice GAPs}\label{sec:choicegaps}
In this section, we formally define a social network (SN) and introduce the \emph{Choice Generalized Annotated Program (\cGAP)} paradigm.

\subsection{Social Network Formalization}

We assume the existence of two arbitrary but fixed disjoint sets $VP, EP$ of
\emph{vertex} and \emph{edge predicate symbols} respectively. Each vertex
predicate symbol has arity 1 and each edge predicate symbol has arity 2.

\begin{definition}
 A \emph{social network} is a 5-tuple $(\V, \E, \la_{vert}, \la_{edge} , w)$
where:
\begin{enumerate}
\item $\V$ is a finite set whose elements are called \emph{vertices}.
\item $\E \subseteq \V \times \V$ is a finite multi-set whose elements are
called edges.
\item $\la_{vert} : \V \rightarrow 2^{VP}$ is a function, called the vertex labeling
function.
\item $\la_{edge} : \E \rightarrow EP$ is a function, called the edge labeling
function.
\item $w : E \rightarrow [0, 1]$ is a function, called a weight function. \hfill$\Box$
\end{enumerate}
\end{definition}

\subsection{Syntax of Choice GAPs (\cGAPs)}
A \cGAP\ consists of two parts:

\begin{enumerate}
\item An ``annotation'' language;
\item A logical language that is connected to the annotation language via certain shared syntactic elements.
\end{enumerate}

\subsubsection{\cGAP\ Annotation Language}
Let $AVar$ be a set of symbols (called ``annotated variable symbols'') ranging over the unit real interval $[0,1]$
and let ${\cal F}$ be a set of symbols (called ``annotation function symbols''), each with an associated arity.

\begin{definition}[Annotation]
Annotations are inductively defined as follows:
 \begin{enumerate}
  \item Any member of $[0,1]\cup AVar$ is an annotation.
  \item If $f \in \cal F $ is an $n$-ary annotation function symbol and $t_1,\dots,t_n$ are annotations, then $f(t_1,\dots,t_n)$ is an annotation.\hfill$\Box$
 \end{enumerate}
 \end{definition}
As in the case of Generalized Annotated Programs \cite{ks92}, note that each annotation function symbol $f$ of arity $i$ denotes some fixed pre-theoretically defined function from $[0,1]^i$ to $[0,1]$.

\subsubsection{\cGAP\ Logical Language}
We define a separate  logical language whose constants are members of $\V$ and whose predicate
symbols consist of $VP \cup EP$. We also assume
the existence of a set $\cal V$ of variables ranging over the constants (vertices).
No function symbols are present.
Terms and atoms are defined in the usual way (cf. [Lloyd 1987]).
If $A = p(t_1 ,\dots,t_n )$ is an atom and $p \in VP$ (resp. $p \in EP$), then $A$ is called a vertex (resp. edge) atom.
%We will use $\cal A$ to denote the set of all ground atoms (i.e. atoms where no variable occurs).

\begin{definition}[Annotated atom/ GAP-rule/GAP]
If $A$ is an atom and $\mu$ is an annotation, then $A:\mu$ is an \emph{annotated
atom}.
If $A$ is a vertex (resp. edge) atom, then $A:\mu$ is also called a vertex (resp.
edge)
annotated atom.
If $A_0:f(\mu_1,\dots,\mu_n)$, $A_1:\mu_1,\dots,A_n:\mu_n$ are annotated atoms,
then
$$A_0:f(\mu_1,\dots,\mu_n) \leftarrow A_1:\mu_1,\dots,A_n:\mu_n$$
is an \emph{annotated rule}. When $n=0$, the above rule is called a \emph{fact}.
A \emph{generalized annotated program (GAP)} is a finite set of annotated rules.
An annotated atom (resp. a rule, a GAP) is \emph{ground} iff there are no
occurrence of variables
from either $\AVar$ or $\Var$ in it.\hfill$\Box$
\end{definition}

Every social network $SN = (\V, \E, \la_{vert}, \la_{edge}, w)$ can be represented by the set of GAP-rules (actually facts)
$\Pi_{SN} = \{q(v) : 1 \leftarrow |\ v\in  \V \wedge q \in \la_{vert} (v)\} \cup
\{ep(v_1 , v_2) : w( \langle v_1 , v_2 \rangle) \leftarrow |\ \langle v_1 , v_2 \rangle \in \E  \wedge \la_{edge}(\langle v_1 , v_2 \rangle) = ep\}$.
To construct a GAP from a social network $SN$, we look at each vertex $v$ and each property $q$. If $v$ has property $q$, then $q(v):1\leftarrow$ is inserted into $\Pi_{SN}$.
Likewise, we look at each edge $(v_1,v_2)\in \E$. If this edge has weight $w$ and edge property $ep$, then we insert the fact $ep(v_1,v_2):w\leftarrow$ into $\Pi_{SN}$. \cite{SNDOP_query} has already shown that many classical diffusion models such as
the SIR and SIS models for disease spread~\cite{anderson79,Hethcote76}, the Jackson-Yariv model for diffusion of product adoption~\cite{jy05}, the Flickr diffusion model \cite{ChaMisloveGummadi_2009}, and homophilic models \cite{aral09} can all be expressed as GAPs.

\cGAPs\ extend GAPs by adding a single rule called a \emph{Vertex Choice (VC) Rule} inspired by Sacc{\`a} and Zaniolo\cite{SacZan90} for classical Datalog. Every \cGAP\ consists of a GAP together with a single vertex choice rule.

\begin{definition}[Vertex choice (VC) rule]
Suppose $\{a_1,\dots,a_m\}$ and
$\{b_1,\dots,b_m\}$ are two ordered sets of vertex predicate symbols. Then
$$b_1(X),\dots,b_m(X) \hookleftarrow a_1(X),\dots,a_m(X)$$
is a \emph{vertex choice (VC) rule} of size $m$ for the vertex $X$. A VC-rule is ground iff there are no occurrence of variables from $\Var$ in it.
 \hfill$\Box$
\end{definition}
Note that edge predicate symbols cannot appear anywhere inside a VC-rule. Moreover, usually, only conflicting predicate symbols occur within a VC-rule and usually the predicate symbol
$b_i$ is the decision predicate corresponding to a utility predicate $a_i$. VC-rules do not contain any annotations.

For instance, 
\begin{eqnarray*}
voteTory^D(X), voteLabour^D(X) & \hookleftarrow & voteTory^U(X), voteLabour^U(X)
\end{eqnarray*}
is a VC-rule from our earlier example. Intuitively, this rule says that if $voteTory^U(X),$ $voteLabour^U(X)$ are true, then exactly one of $voteTory^D(X),voteLabour^D(X) $ can be true and that a choice must be made.

\begin{definition}[Choice GAP]
A \emph{Choice GAP} (\cGAP)  $\Pi$ is a finite set of annotated
rules plus a single vertex choice rule.
\end{definition}
\emph{Though choice rules have only been applied in classical Datalog settings in the past and their combination with GAPs is new, we do not claim this as a major contribution of this paper as the combination is straightforward. However, the use of \choicegaps\ for modeling competitive diffusion, and the new notions of equilibria we introduce, together with appropriate complexity results, algorithms, and efficiently computable fragments of \choicegaps\ are new.}

\subsection{Semantics of \cGAP}
We are now ready to define the semantics of \cGAPs.
Given a \cGAP\ $\Pi$, let $\atoms$ denote the set of all ground atoms of $\Pi$.

\begin{definition}[Interpretation]
Given a a \cGAP\ $\Pi$, an interpretation $I$ for $\Pi$ is any mapping $I: \atoms
\rightarrow [0,1]$ of ground atoms to real numbers in $[0,1]$. \hfill$\Box$
\end{definition}
Thus, an interpretation merely assigns a certainty value to each ground atom in $\atoms$.
The set ${\cal I}$ of all interpretations can be partially ordered via the ordering $\preceq$ defined as follows:
$I_1 \preceq I_2$ iff for
all ground atoms $A$, $I_1(A) \leq I_2(A)$. ${\cal I}$ forms a complete lattice under the $\preceq$ ordering.
Given two interpretations $I_1$ and $I_2$, we define their intersection $I_1 \cap I_2$ as the interpretation
$(I_1\,\cap\, I_2)$ such that $(I_1\,\cap\, I_2)(A) = \min(I_1(A),I_2(A))$ for all $A\in\atoms$.
Similarly, the union $I_1 \cup I_2$ of interpretations $I_1$ and $I_2$ is the interpretation $I_1\,\cup\, I_2$ such that $(I_1\,\cup\, I_2)(A)=\max(I_1(A),I_2(A))$ for all $A\in\atoms$.
We are now ready to define satisfaction.

\begin{definition}[Satisfaction]
Let $I$ be an interpretation.
\begin{itemize}
 \item $I$ satisfies a ground annotated atom $A:\mu$, denoted $I \models A:\mu$,
iff $I(A) \geq \mu$.
 \item $I$ satisfies a ground \cGAP \ annotated rule $r$ of the form
       $A_0:\mu_0 \leftarrow A_1:\mu_1,\dots,A_n:\mu_n$, denoted $I \models r$,
       iff $I(A_0) \geq \mu_0$ or for some $i \in \{1,\dots,n\}$, $I \not\models
A_i:\mu_i$.
 \item $I$ satisfies a ground VC-rule $r$ of the form
       $B_1,\dots,B_m \hookleftarrow A_1,\dots,A_m$, denoted $I \models
r$, iff
       $\exists i \in \{1,\dots,m\}$ such that $I(B_i) = I(A_i)$ and $\forall j
\in \{1,\dots,m\}$,
       $j \neq i$, $I(B_j) = 0=I(A_j)$. 
\item $I$ satisfies a non-ground GAP/VC rule iff it satisfies all ground instances of it.
\item $I$ satisfies a \cGAP\ $\Pi$ (or is a \emph{model} of $\Pi$) iff $I$ satisfies all rules in $\Pi$.
\hfill$\Box$
\end{itemize}
\end{definition}
A key part of this definition is the satisfaction of VC-rules. For $I$ to satisfy a VC-rule $r$ of the form shown above, there must exist exactly one pair $(A_i,B_i)$ such that $I(A_i)=I(B_i)\geq 0$. For all other pairs
$(A_j,B_j)$, we have $I(A_j)=I(B_j)=0$.
We now provide a simple example.

\begin{example}\label{es:inter}\em
Consider a social network $SN$ and two diffusion models $DM_1$
and $DM_2$ relating to diffusion about the tendency to buy ASUS computers versus buying Macs. For this toy example which will be used throughout the paper, we assume these are the only two options of computers to buy (the same reasoning works if there are $n$ different computers to buy). We assume there are two vertices $1,2$ and a friend edge from $1$ to $2$.

$$SN =
\begin{array}{lcl}
 friend(1,2):1 & \leftarrow &
\end{array}
$$

$$DM_1 = \left\{
\begin{array}{lcl}
buyAsus^U(1):0.6 & \leftarrow &\\
buyAsus^U(Y):\mu & \leftarrow & friend(X,Y):1, buyAsus^D(X):\mu
\end{array} \right.
$$

$$DM_2 = \left\{
\begin{array}{lcl}
buyMac^U(1):0.3 & \leftarrow &\\
buyMac^U(Y):\mu & \leftarrow & friend(X,Y):1, buyMac^D(X):\mu
\end{array} \right.
$$

Suppose we have the vertex choice rule $$r: buyMac^D(X),buyAsus^D(X)
\hookleftarrow buyMac^U(X),buyAsus^U(X)$$
Consider the two interpretations $I_1$ and $I_2$ shown below.

\begin{center} \scriptsize
\begin{tabular}{|c|c|c|c|c|}
\hline
&$buyAsus^U(1)$ & $buyAsus^D(1)$ & $buyAsus^U(2)$ & $buyAsus^D(2)$ \\ \hline
$I_1$ &$0.6$ &$0.6$ & $0.6$ & $0.6$\\
$I_2$ &$0.6$ & $0.6$ & $0.6$ & $0.0$ \\ 
\hline
 &$buyMac^U(1)$ & $buyMac^D(1)$ &
$buyMac^U(2)$ & $buyMac^D(2)$\\ \hline
$I_1$ & $0.3$ & $0.0$ & $0.3$ & $0.0$\\ 
$I_2$  & $0.3$ & $0.0$ & $0.7$ & $0.7$\\ \hline
\end{tabular}
\end{center}

Consider the situation of vertex $1$. Interpretation $I_1$ assigns:

\begin{itemize}
\item $0.6$ to all  ground atoms $buyAsus^D(1), buyAsus^U(1), buyAsus^D(2),buyAsus^U(2)$, 
\item $0.3$ to $buyMac^D(1)$, 
\item $0$ to $buyMac^U(1)$, 
\item $0.3$ to
$buyMac^U(2)$, and 
\item $0$ to $buyMac^U(2)$.
\end{itemize}
We see that $I_1$ satisfies all the diffusion rules as well as the one VC-rule. Consider the ground instance of this VC-rule with $X=1$. Exactly one of the two head decision atoms, $buyAsus^D(1)$ has a value greater than $0$ (0.6) and this coincides with the value assigned by $I_1$ to $buyAsus^U(1)$.
Likewise, when we consider the ground instance with $X=2$, we see the same thing. Thus, $I_1$ is a model of the
\cGAP \ program
$\Pi = SN \cup DM_1 \cup DM_2 \cup \{r\}$.

In the same way, we can also establish that $I_2$ is also a model of $\Pi$.
\hfill$\Box$
\end{example}

\section{Coherent Models and Strong Equilibria}
The new technical contributions of this paper start here.

Though $I_2$ is a model of $\Pi$ in the above example, it assigns overly high utilities.
For instance, consider the second rule of $DM_2$ with the substitution $\theta=\{X=1,Y=2\}$. We know that for the ground atom $buyMac^D(1)$ in the body of this rule after $\theta$ is applied to it, $I_2(buyMac^D(1))=0$.
But the head of this rule under substitution $\theta$, which is the atom $buyMac^U(2)$ is assigned a utility of $0.7$ instead of the $0$ that is the minimum needed for this rule to be satisfied. 
In order to address this, we define the concept of a \emph{coherent} model.

\begin{definition}[Coherence Transformation]\label{def:coh-transform}
Suppose $\Pi$ is a \cGAP, $r\in ground(\Pi)$ is an instance of the single VC-rule in $\Pi$, and $I$ an interpretation. Suppose $r$ has the form $B_1(v),\dots,B_m(v) \hookleftarrow A_1(v),\dots,A_m(v)$
The \emph{coherence-transform of $r$} is the set $\coh(r,I)= \{ B_i(v):\mu \leftarrow A_i(v):\mu\: |\: I(A_i(v)) > 0$ and
$I(A_i(v)) = I(B_i(v))\}$. Note that this set can be empty.

The \emph{coherence transform} of $\Pi$ w.r.t. $I$ is simply the GAP $ground(\Pi^{non\_vc})\,\cup\, \bigcup_{\scriptsize \begin{array}{l}
r\in ground(\Pi) \wedge\\
r\ \mbox{is a VC-rule}
\end{array}} \coh(r,I)$, where $\Pi^{non\_vc}$ is the set of all non-VC rules in $\Pi$.
\end{definition}
Thus, the coherence transform of $\Pi$ w.r.t. $I$ simply looks at ground VC-rules in $ground(\Pi)$. If there is a ground atom $A_i(v)$ in the body of the rule such that $I(A_i(v)) > 0$ and $I(A_i(v))=I(B_i(v))$, then we include the GAP rule $B_i(v):I(B_i(v))\leftarrow A_i(v):I(B_i(v))$ in $\coh(\Pi,I)$ --- otherwise we just get rid of the rule. All non-VC rules of $\Pi$ are included in $\coh(\Pi,I)$. \emph{Thus, $\coh(\Pi,I)$ is always a GAP which, by \cite{ks92}, is guaranteed to have a unique minimal model.} We use ${\cal MM}(\Pi)$ to denote the minimum model of a GAP $\Pi$.  
We can now define coherent models.

\begin{definition}[Coherent model]\label{def:stableModel}
Let $\Pi$ be a \cGAP\ and let $M$ be a model for $\Pi$.
$M$ is a \emph{coherent model} for $\Pi$ iff it is the minimum model of the
GAP $\coh(\Pi,M)$.
\hfill$\Box$
\end{definition}
We now present a quick example of coherent models.

\begin{example}\label{es:coemodel}\em
We show that the model $I_1$ from Example~\ref{es:inter} is a coherent model for the \cGAP\  $\Pi$ below.
$$\Pi = \left\{
\begin{array}{lcl}
friend(1,2):1 & \leftarrow & \\
buyAsus^U(1):0.6 & \leftarrow &\\
buyAsus^U(Y):\mu & \leftarrow & friend(X,Y):1, buyAsus^D(X):\mu \\
buyMac^U(1):0.3 & \leftarrow &\\
buyMac^U(Y):\mu & \leftarrow & friend(X,Y):1, buyMac^D(X):\mu \\
buyMac^D(X),buyAsus^D(X) & \hookleftarrow & buyMac^U(X),buyAsus^U(X)
\end{array}\right.
$$
Let $r$ be the single VC-rule in $\Pi$.
By grounding $\Pi$ we obtain
$$ground(\Pi) = \left\{
\begin{array}{lcl}
friend(1,2):1 & \leftarrow & \\
buyAsus^U(1):0.6 & \leftarrow &\\
buyAsus^U(2):0.6 & \leftarrow & friend(1,2):1, buyAsus^D(1):0.6 \\
buyMac^U(1):0.3 & \leftarrow &\\
%buyMac^U(Y):\mu_2 & \leftarrow & friend(1,2):1, buyMac^D(1):0 \\
buyMac^D(1),buyAsus^D(1) & \hookleftarrow & buyMac^U(1),buyAsus^U(1)\\
buyMac^D(2),buyAsus^D(2) & \hookleftarrow & buyMac^U(2),buyAsus^U(2)
\end{array}\right.
$$

Consider each of the two ground VC-rules above. For the first VC-rule, we see that $I_1(buyMac^D(1)) \neq I_1(buyMac^U(1))$ and $I_1(buyAsus^D(1)) = I_1(buyAsus^U(1)) $ $= 0.6 > 0$.
Hence, the rule
\begin{eqnarray*}
buyAsus^D(1):0.6 & \leftarrow & buyAsus^U(1):0.6
\end{eqnarray*}
belongs to the set $\coh(r,I_1)$, and gets added to the coherent transform of $\Pi$ w.r.t. $I_1$, denoted by $\Pi'$ in the following. Likewise, with the second VC-rule, we know that 
$I_1(buyMac^D(2)) = I_1(buyMac^U(2)) = 0$,
and $I_1(buyAsus^D(2)) = I_1(buyAsus^U(2)) $ $= 0.6 > 0$ and so we add the following GAP rule to $\Pi'$:
\begin{eqnarray*}
buyAsus^D(2):0.6 & \leftarrow & buyAsus^U(2):0.6.
\end{eqnarray*}
The final GAP  $\Pi'$ that we get is the following:

$$\Pi' = \left\{
\begin{array}{lcl}
friend(1,2):1 & \leftarrow & \\
buyAsus^U(1):0.6 & \leftarrow &\\
buyAsus^U(2):0.6 & \leftarrow & friend(1,2):1, buyAsus^D(1):0.6 \\
buyMac^U(1):0.3 & \leftarrow &\\
%buyMac^U(Y):\mu_2 & \leftarrow & friend(1,2):1, buyMac^D(1):0 \\
buyAsus^D(1):0.6 & \leftarrow & buyAsus^U(1):0.6\\
buyAsus^D(2):0.6 & \leftarrow & buyAsus^U(2):0.6
\end{array} \right.
$$

It is easy to see that $I_1$ is the minimal model of $\Pi'$. Hence, $I_1$ is a coherent model of $\Pi$. \hfill$\Box$
\end{example}

 Algorithm~\ref{algo:allStable} presents a non-deterministic naive algorithm to compute all coherent models of $\Pi$. The algorithm looks at each VC-rule and non-deterministically chooses a pair $(A_i(v),B_i(v))$. If we think of these as the appropriate pair of atoms from the VC-rule, we insert the annotated rule
$B_i(v):Y\leftarrow A_i(v):Y$. Such a rule is added for each ground VC-rule instance and we then compute the minimal model of the resulting GAP. If this minimal model satisfies every VC-rule, then it is clearly a coherent model --- otherwise it is not.

\begin{algorithm}[t]
\caption{Non-deterministic algorithm enumerating all coherent models.}\label{algo:allStable}
\begin{algorithmic}[1]
\Procedure{$coherentModel$}{\cGAP\ program $\Pi$, $m$ competing choices}
\State Let $\Pi' = ground(\Pi)$
\State $X=\{ r\: |\: r\in ground(\Pi)$ and $r$ is a VC-rule $\}$. 
\State $\Pi'=\Pi'-X$.
\For{each vertex choice rule $B_1(v),\dots,B_m(v) \hookleftarrow A_1(v),\dots,A_m(v)$ in $X$}
\State guess a choice $i \in \{1,2,\dots,m\}$
\State insert the rule $B_i(v):\mu \leftarrow A_i(v):\mu$ into $\Pi'$
\EndFor
\State Let $I$ be the minimum model of $\Pi'$
\If{$I$ satisfies every vertex choice rule in $X$}
\State \Return $I$
\Else\ \ \emph{fail}
\EndIf \EndProcedure
\end{algorithmic}
\end{algorithm}

%\begin{proposition}
%Algorithm~\ref{algo:allStable} returns the set of all coherent models.
%\end{proposition}

The following result shows that the $coherentModel$ function correctly finds the set of all coherent models.

\begin{rproposition}[\propalgo]
Algorithm~\ref{algo:allStable} returns the set of all coherent models.
\end{rproposition}

We now introduce the concept of \SE.

\begin{definition}[Strong Equilibrium]
A coherent model $I$ is a \emph{\SE} for a \cGAP \ $\Pi$ iff
%$I$ satisfies all \cGAP \ rules in $\Pi$ and
for each ground vertex choice rule
of the form $B_1,\dots,B_m \hookleftarrow A_1,\dots,A_m$ it is the case that
% \begin{eqnarray*}
% \sum_{i=1}^m I(B_i) & \geq & I(A_1)\\
%  & \vdots &\\
% \sum_{i=1}^m I(B_i) & \geq & I(A_m)
% \end{eqnarray*}
$$\sum_{i=1}^m I(B_i) = \max(I(A_1),\dots,I(A_m))$$
\hfill$\Box$
\end{definition}

%\from{vs}{all}{explain why this is called an SE}

Recall that by the definition of VC-rule satisfaction, there exists only one $B_i$ such that $I(B_i) \geq 0$, while, for all other $B_j$, for $j \neq i$, $I(B_j) = 0$.
Thus, a \SE\ is a coherent model where each choice coincides with the maximum annotation value, taken as a measure of utility, in the VC-rule body.
We use $\eqs(\Pi)$ to denote the set of all \SEs\
 of a \cGAP\ $\Pi$. The set of all \SEs\ can be computed by enumerating all coherent models using Algorithm~\ref{algo:allStable}, and then selecting those models that are also \SEs.

\begin{example} The coherent model $I_1$  from  Examples~\ref{es:inter}~and~\ref{es:coemodel} is a \SE\ because  $$I_1(buyAsus^D(1))+I_1(buyMac^D(1))=0.6+0.0=$$ $$=\max(I_1(buyAsus^U(1)),I_1(buyMac^U(1)))=\max(0.6,0.3)=0.6$$ and $$I_1(buyAsus^D(2))+I_1(buyMac^D(2))=0.6+0.0=$$ $$=\max(I_1(buyAsus^U(2)),I_1(buyMac^U(2)))=\max(0.6,0.3)=0.6$$~\hfill~$\Box$
\end{example}

We are now ready to define when a $\cGAP$ entails an annotated atom.

\begin{definition}[Entailment]
 A $\cGAP$  $\Pi$ entails a ground annotated atom $AA$, denoted $\Pi \models AA$,
 iff every \SE\ of $\Pi$ satisfies $AA$.\hfill$\Box$
\end{definition}

\section{\cGAPs\ : A  Game Perspective}\label{sec:game-formalization}
Let $\Pi$ be a \cGAP,  SN be a social network, and let $n$ be the number of vertices in SN. %, and let $m$ be the number of vertex choice rules in $\Pi$.
Each vertex $v$ is considered to be a player $\calp_v$.
We use $\Gamma_{\Pi}$ to denote the set of all players in $\Pi$. In this section, we first describe the concept of a state (which basically is an a mapping of players to actions, specifying the action the player takes). We develop a formal definition of a Nash equilibrium for the resulting game, as well as a relationship between states and strong equilibria for the game.  This
 allows us to go back and forth between states that ``represent'' a strong equilibrium and strong equilibria.

Each player $\calpv\in \Gamma_{\Pi}$ has a the same set of actions (or strategies) $Q=\{1,\dots,m\}$ where $m$ is the size of the vertex choice rule in $\Pi$. These are the $m$ competing choices the player can make
(e.g. buying an Asus vs. buying a Mac, or voting Tory vs. voting for Labour).

A \emph{state} $S$ for a \cGAP\ $\Pi$ represents a choice for each player $\calpv$ and it is defined as a mapping
$S: \Gamma_{\Pi} \rightarrow Q$.

Given a \cGAP \ $\Pi$ and a state $S$ for $\Pi$, we define the notion of an \emph{induced ground GAP} $\Pi_S$.

\begin{definition}[Induced Ground GAP $\Pi_S$]\label{def:inducedGAP}
Suppose $\Pi$ is a \cGAP\ and $S$ is a state for $\Pi$. We define a GAP $\Pi_S$ that can be obtained from
$\Pi$ and $S$ as follows:

\begin{enumerate}
\item
replace each ground VC-rule $r:b_1(v),\dots,b_m(v) \hookleftarrow a_1(v),\dots,a_m(v)$ in $ground$ $(\Pi)$,
with the ground annotated rule
$b_i(v):X \leftarrow a_i(v):X$ where $i=S(\calpv)$. 
\item All non-VC rules in $\Pi$ are also in $\Pi_S$.
\end{enumerate}
\end{definition}
Intuitively, when considering the ground instance  $r$ of a VC-rule in $\Pi$ and a state $S$, exactly one of the $b_i(v)$'s can be true in the state as a vertex $v$ can make exactly one choice. This choice is
the $i=S(\calpv)$. 

\begin{rproposition}[\statecoe]\label{statecoe}
Let $\Pi$ be a \cGAP such that every predicate appearing in the head of the VC-rule does not
appear in the head of a GAP rule, and let $S$ be a state. Then, the minimal model of $\Pi_S$ is a coherent model for $\Pi$.
\end{rproposition}

Given a state $S$, each player has a \emph{utility value} for each of its actions.
The utility $u(S,\calpv,i)$ of the player $\calpv$  performing the action $i \in Q$ in the state $S$ is given by the value assumed by the atom $a_i(v)$ in the interpretation ${\cal MM}(\Pi_S)$, i.e. we set
$u(S,\calpv,i) = {\cal MM}(\Pi_S)(a_i(v))$, where $a_i(x)$ is the $i$'th atom in the body of the VC-rule. This value is the likelihood of the player performing action $i\in Q$ according to the GAP $\Pi_S$.

We assume that each player is a rational agent, i.e. he is motivated by maximizing his own payoff.

\begin{definition}[state representation of strong equilibria]
A state $S$ \emph{represents a \SE} for $\Pi$ iff for all players $\calpv \in \Gamma_{\Pi}$
$$u(S,\calpv,S(\calpv)) \geq u(S,\calpv,i)$$
for each action $i \in Q$.
\end{definition}

Intuitively, a state is a choice of actions, one for each player. In contrast, strong equilibria, as defined in the previous section, is a coherent model of a \cGAP\ that satisfies certain equilibrium conditions. The above definition specifies the relationship between states and strong equilibria so we can refer to the actions taken in a strong equilibrium as a state and vice versa.

\begin{fact}
 The set $\{ \mathcal{MM}(\Pi_S)\: |\: S $ is a state $\}$, contains all \SEs\ for $\Pi$.\hfill$\Box$
\end{fact}

The above result specifies a set containing all possible strong equilibria - but not all members of  $\{ \mathcal{MM}(\Pi_S)\: |\: S $ is a state $\}$ are necessarily strong equilibria.

\subsection{Nash equilibrium vs. \SE}
%(TO REWRITE WITH THE CONCEPTS OF PERFECT KNOWLEDGE AND RATIONALITY)
A Nash equilibrium is a state where no player has anything to gain by unilaterally changing his own action. In order to define Nash equilibria, we first define the utility $\hat{u}(S,\calpv)$ of a state $S$ for a player $\calpv$
as follows:
$$\hat{u}(S,\calpv) = u(S,\calpv,S(\calpv)).$$
This definition says the utility of the state $S$ for player $\calpv$ is simply the utility of the action $S(\calpv)$ that he takes in that state.

\begin{definition}[Nash equilibrium]
Let $\Pi$ be a \cGAP\ and $S$ a state.
${\cal MM}(\Pi_S)$ is a Nash equilibrium for $\Pi$
iff for each player $\calpv$
$$\hat{u}(S,\calpv) \geq \hat{u}(S',\calpv)$$
for each $S'$ such that $S(\calp_{v'}) = S'(\calp_{v'})$ if $v' \neq v$, and $S(\calp_{v'}) \neq S'(\calp_{v'})$ if $v'= v$.\hfill$\Box$
\end{definition}
Intuitively, ${\cal MM}(\Pi_S)$ is a Nash equilibrium if there is no other state $S'$ in which all the players can obtain a higher utility. Thus, if one player tries to perform an action different from that in a Nash equilibrium (trying to raise his own utility), this would imply a reduced utility for some other player, who may then try to perform some other action, leading to an unstable situation.

The above definition of classical Nash equilibrium applied to competitive diffusion in SNs assumes that
% Then, the main differences between the classical Nash equilibrium and the \SE\ are that
% the Nash equilibrium assumes that
each player has common knowledge about:
\begin{enumerate}
\item the whole structure of the social network (and every vertex in it),
\item for all players, how they think (diffusion model mechanism for each vertex),
\item the strategies adopted by each other player.
\end{enumerate}

All these assumptions are needed to compute the utility $\hat{u}(S',p)$ --- unfortunately, they are too strong for
a real-world social network context. In most real-world social networks, we have information on our neighbors but not on others. Likewise, we are not privy to the strategies of the players and how they make decisions.
 Fortunately, the important theorem below says that we do not need these assumptions --- strong equilibria can capture Nash equilibria for all games. We now recall the definition of a generic game from  classical game theory.

There is no loss of generality in the assumption that the set of all actions is the same for each player. If  some players have actions that other players cannot perform, then the utility of performing those other actions can be set to 0 for players who cannot perform them.
\begin{definition}[Generic game]
A \emph{generic game} $G$ is a triple $G=(\hat{P},\hat{Q},\hat{U})$ where
\begin{itemize}
 \item $\hat{P}$ is the set of players $\{p_1,\dots,p_n\}$,
 \item $\hat{Q}=\{q_1,\dots,q_m\}$ is the set of actions (the same for each player), and
  \item $\hat{U}=\{\hat{u}_1,\dots,\hat{u}_n\}$ is the set of utility functions, one for each player,
  where a utility function is a function $\hat{u}_i : \underbrace{\hat{Q} \times \dots \times \hat{Q}}_n \rightarrow  \Re$,
\end{itemize}
\end{definition}
We now state our important representation theorem.

\begin{rtheorem}[\nasheqThm]{\scshape Nash equilibria can be captured by strong equilibria of \cGAPs.}
\label{th:SEcapNash}
\textrm{For every generic game $G=(\hat{P},\hat{Q},\hat{U})$, there exists a \cGAP\ $\Pi$ such that the strong equilibria of $\Pi$ coincide with the Nash equilibria of $G$.}
\end{rtheorem}

We now illustrate the above result on the well-known Prisoner's Dilemma problem.

\begin{example}[Prisoner's Dilemma]\label{prisoner}
Consider the well known game of prisoner's dilemma where
the set of players is $\hat{P} = \{A,B\}$, the set of actions is $\hat{Q} = \{c,d\}$ (cooperate, i.e. confess or defect, i.e. don't cooperate),
and the utility functions $\hat{U}$ are represented in the following matrix
\begin{center}
\begin{tabular}{|c|c|c|}
  \hline
    & Prisoner B cooperates & Prisoner B defects \\ \hline
  Prisoner A cooperates & (-6,-6) & (0,-7) \\ \hline
  Prisoner A defects & (-7,0) & (-1,-1) \\\hline
\end{tabular}
\end{center}
For logic program readers, the entry (0,-7) says that Prisoner A's utility when he cooperates and B defects is 0, while prisoner B's utility (when A cooperates and B defects) is -7.
The equivalent \cGAP\ $\Pi^G$ according to the construction in the preceding proof is obtained as follows:
\begin{itemize}
 \item we add the following vertex choice rule
 \begin{eqnarray*}
 c^D(X), d^D(X) & \hookleftarrow & c^U(X), d^U(X)
 \end{eqnarray*}
 \item we add the following facts
 \begin{eqnarray*}
 c^U(A):0.1 \leftarrow && \\
 c^U(B):0.1 \leftarrow && \\
 d^U(A):0.1 \leftarrow && \\
 d^U(B):0.1 \leftarrow &&
 \end{eqnarray*}
 where we used $\varepsilon = 0.1$
 \item by considering the scaled (in the interval $[0,0.9]$) version of the utility matrix
 \begin{center}
\begin{tabular}{|c|c|c|}
  \hline
    & Prisoner B cooperates & 
    Prisoner B defects \\ \hline
 \begin{tabular}{c} Prisoner A\\ cooperates \end{tabular}& ($\mu^A_{(c,c)}=0.13$, $\mu^B_{(c,c)}=0.13$) & ($\mu^A_{(c,d)}=0.9$, $\mu^B_{(c,d)}=0$) \\ \hline
  \begin{tabular}{c}Prisoner A\\ defects \end{tabular}& ($\mu^A_{(d,c)}=0$, $\mu^B_{(d,c)}=0.9$) & ($\mu^A_{(d,d)}=0.78$,$\mu^B_{(d,d)}=0.78$) \\
  \hline
\end{tabular}
\end{center}
we add the following rules
\begin{eqnarray*}
 c^U(A):0.23 & \leftarrow & c^D(B):0.1\\
 c^U(A):1.00 & \leftarrow & d^D(B):0.1\\
 d^U(A):0.10 & \leftarrow & c^D(B):0.1\\
 d^U(A):0.88 & \leftarrow & d^D(B):0.1\\
 c^U(B):0.23 & \leftarrow & c^D(A):0.1\\
 c^U(B):1.00 & \leftarrow & d^D(A):0.1\\
 d^U(B):0.10 & \leftarrow & c^D(A):0.1\\
 d^U(B):0.88 & \leftarrow & d^D(A):0.1
 \end{eqnarray*}
\end{itemize}

The Nash equilibrium for the prisoner's dilemma is the state $(c,c)$, i.e. both prisoners decide to confess.
The minimal model of the program $\Pi^G_{(c,c)}$ induced by the state $(c,c)$  is
${\cal MM}(\Pi^G_{(c,c)}) = \{c^U(A):0.23, c^U(B):0.23,
c^D(A):0.23, c^D(B):0.23, d^U(A):0.10, d^U(B):0.10\}$ which is a \SE\ for $\Pi^G$, because
\begin{eqnarray*}
{\cal MM}(\Pi^G_{(c,c)})(c^D(X)) \geq {\cal MM}(\Pi^G_{(c,c)})(c^U(X)) &&\\
{\cal MM}(\Pi^G_{(c,c)})(c^D(X)) \geq {\cal MM}(\Pi^G_{(c,c)})(d^U(X))
\end{eqnarray*}
for $X \in \{A,B\}$.

In a similar way it is possible to check that the minimal model of the programs induced by the remaining states
are not \SEs\ for $\Pi^G$.\hfill$\Box$
\end{example}

\emph{The concluding message of this section is simple. \cGAPs\ and strong equilibria together can express the notion of Nash equilibria in general games. As a consequence, in the rest of this paper, we will consider
strong equilibria of \cGAPs.}

The following example shows a \cGAP\ where there exists a strong equilibrium that is not a Nash equilibrium.

\begin{example}\label{ex:2eq}
Consider the following program
$$
\begin{array}{lcl}
buyMac^U(1):0.3 & \leftarrow & \\
buyAsus^U(1):0.3 & \leftarrow &\\
buyAsus^U(1):0.6 & \leftarrow &buyAsus^D(1):0.3\\
buyMac^D(X),buyAsus^D(X) & \hookleftarrow & buyMac^U(X),buyAsus^U(X)
\end{array}$$

We have only two \SEs, one for each choice of player $1$

\begin{center} \scriptsize
\begin{tabular}{|c|c|c|c|c|}
\hline
&$buyAsus^U(1)$ & $buyAsus^D(1)$ & $buyMac^U(1)$ & $buyMac^D(1)$  \\ \hline
$SE_a$ &$0.6$ &$0.6$ & $0.3$ & $0.0$\\
$SE_b$ &$0.3$ & $0.0$ & $0.3$ & $0.3$ \\ 
\hline
\end{tabular}
\end{center}

The strong equilibrium $SE_b$ is not a Nash equilibrium because if player $1$ changes his choice ($buyMac$)
to $buyAsus$, his utility increases (see strong equilibrium $SE_a$).~\hfill$\Box$
\end{example}

\subsection{Social Network Game and Obligatory Product Selection}
\cite{gandalf13} define a social network game where all users of the social network must choose one product from among a set of products and their utilities depend on the choices of their neighbors. This paper defines a social network as follows.

\begin{definition}\cite{gandalf13}
A social network consists of:
\begin{itemize}
\item a weighted  directed graph $G_{sn}=(V,E,W)$ where 
\begin{itemize}
\item  $V$ is the set of agents,
\item  $E$ is a set of directed edges between agents,
\item $W:E\rightarrow (0,1]$ is a weighted function associating a weight with each edge;
\end{itemize}
\item a set ${\cal P}=\{1,\dots, h\}$ of products;
\item a function $P:V\rightarrow 2^{P}\setminus \emptyset$ associating, with each agent, a non empty subset of products;
\item a threshold function $\theta:V\times P\rightarrow (0,1]$ associating, with each agent and each product, a threshold value.\hfill$\Box$
\end{itemize}
\end{definition}

Each agent $v\in V$ chooses a product in $P(v)$. Let $s$ denote the joint strategy of the agents, i.e.
$s:V\rightarrow {\cal P}$ assigns a product to each vertex.

Let $c_0$ be a constant in $(0,1]$ and $N(i)=\{j|(j,i)\in E\}$. 
The utility function $\overline{u}_v$, 
for each agent $v\in V$ and for each joint strategy $s$, is defined as follows:

$$\overline{u}_v(s) = \begin{cases}
c_0 & if \:\:N(v)=\emptyset\\
1/2+\frac{1}{2*|N(v)|}\sum_{i\in N(v),s_i=s_v} w((i,v))-\theta(v,s_i) & otherwise
\end{cases}$$

Note that in the utility, we add the term $1/2+\frac{1}{2*|N(v)|}$ in order to ensure that the utility function returns a value in the $[0,1]$ interval. Obviously, this modification does not change the game. 
We call this game, the Apt-Simon Social Network Game and denote its by the tuple $(G_{sn},{\cal P},P,\theta,c_0)$.

\begin{rtheorem}[\thsng]{\scshape Apt-Simon Social network game and strong equilibria of \cGAPs.}\label{thsng}
For every social network game $(G_{sn},{\cal P},P,\theta,c_0)$ there exists a \cGAP\ $\Pi$ such that the strong equilibria of $\Pi$ coincide with the Nash equilibria of $(G_{sn},{\cal P},P,\theta,c_0)$.
\end{rtheorem}

The following theorem provides a new special case of Apt-Simon social network games where a Nash equilibrium always exists and can be computed in PTIME.

\begin{rtheorem}[\thmSEcoincidesNE]
For every social network game $(G_{sn},{\cal P},P,\theta,c_0)$ where ${\cal P}$ has only two products ($|{\cal P}|=2$) a Nash equilibrium always exists and can be computed in PTIME.
\end{rtheorem}

\section{\cGAP\ complexity}\label{sec:complexity}
In this section, we study the computational complexity of various problems related to strong equilibria.
We start by  noting that 
strong equilibria may not always exist. This is due to two factors,
\begin{itemize}
 \item there exist rule heads (or facts) containing a predicate symbol present in the head of the vertex choice rule, or
 \item atoms expressed in the vertex choice rule have mutual dependencies encoded within the GAP rules, leading to potential ``conflict'' and non-existence of strong equilibria.
\end{itemize}
Examples~\ref{ex:const} and \ref{ex:inter} below show the first and the second case, respectively.

\begin{example}\em \label{ex:const}
Both the programs $\Pi_1$ and $\Pi_2$ do not have any \SE.
%The program $\Pi_1$ does not have any equilibrium model.
$$\Pi_1 = \left\{
\begin{array}{lcl}
buyMac^D(tom):1 & \leftarrow & \\
buyAsus^D(tom):1 & \leftarrow & \\
buyMac^D(X),buyAsus^D(X) & \hookleftarrow & buyMac^U(X),buyAsus^U(X)
\end{array} \right.
$$

%Instead the program $\Pi_2$ has an equilibrium model, but does not have any stable
%equilibrium model.
$$\Pi_2 = \left\{
\begin{array}{lcl}
buyMac^D(tom):1 & \leftarrow & \\
buyAsus^U(tom):1 & \leftarrow & \\
buyMac^D(X),buyAsus^D(X) & \hookleftarrow & buyMac^U(X),buyAsus^U(X)
\end{array}\right.
$$
% In fact the interpretation $I$ s.t. $I(buyMac^D(tom))=1, I(buyAsus^U(tom))=1,
% I(buyMac^U(tom))=1, I(buyAsus^D(tom))=0$
% is the equilibrium model for $\Pi_2$, but it is not stable as is not possible to
% infer $buyMac^U(tom):1$ from $buyMac^U(tom):0$.

In the case of $\Pi_1$, the VC-rule is not satisfied because of
the presence of the two facts containing predicates involved in the head of the
vertex choice rule, while in $\Pi_2$ the vertex choice rule is satisfied.\hfill$\Box$
\end{example}

\begin{example}\em \label{ex:inter} Consider the following program
$$\Pi = \left\{
\begin{array}{lcl}
friend(tom,bob):1 & \leftarrow & \\
buyAsus^U(bob):0.5 & \leftarrow & \\
buyMac^U(X):1 & \leftarrow & friend(X,Y):1, buyMac^D(Y):0.5\\
buyAsus^U(X):1 & \leftarrow & friend(X,Y):1, buyAsus^D(Y):0.5\\
buyMac^U(Y):\mu & \leftarrow & friend(X,Y):1, buyAsus^D(X):\mu\\
buyAsus^U(Y):\mu & \leftarrow & friend(X,Y):1, buyMac^D(X):\mu\\
buyMac^D(X),buyAsus^D(X) & \hookleftarrow & buyMac^U(X),buyAsus^U(X)
\end{array}\right.
$$
Since the vertex predicate $buyMac^U$ depends on $buyAsus^U$ and vice versa, 
a \SE\ does not exist. \hfill$\Box$
\end{example}
Our first complexity result shows that determining existence of strong equilibria is an NP-complete problem.

\begin{rtheorem}[\theqNPco][Strong equilibria existence complexity]\label{th:eqNP-co}
Given a \cGAP \  $\Pi$ as input, the problem of deciding whether there $\Pi$ has a \SE\ is {\bf NP}-complete under data and combined complexity.\hfill$\Box$ %(i.e in the size of the embedded Social Network).
\end{rtheorem}

A major problem occurs when multiple strong equilibria exist. In this case, a player who computes all of these strong equilibria may not know which strong equilibria the other players might act in accordance with.
Thus, he may wish to know if a particular action he is considering is true in all strong equilibria. This problem too is intractable.

\begin{rtheorem}[\thmEnt][Complexity of truth in all strong equilibria] \label{thmEnt}
Given a\\\cGAP \ program $\Pi$ and a ground annotated atom $AA$ as input, 
the problem of deciding whether $\Pi \models AA$ is {\bf coNP}-complete under data and combined complexity.\hfill$\Box$ %(i.e in the size of the embedded Social Network).
\end{rtheorem}

Finally, we show results on the counting complexity of strong equilibria. Not surprisingly, these results also indicate intractability.

\begin{rtheorem}[\countingSE][Counting Complexity of \SEs]\
\begin{itemize}
 \item Given a \cGAP \ program $\Pi$ as input, the problem of counting the number of \SEs\ is {\bf \#P}-complete under data and combined complexity.
 \item Given a positive integer $k$ and \cGAP\ program $\Pi$ as input, the problem of deciding whether $\Pi$ has at least $k$ \SEs\ is in {\bf PSPACE} and {\bf PP}-hard.\hfill$\Box$
\end{itemize}
\end{rtheorem}

\section{MILP \SE\ formulation for linear \cGAPs}\label{sec:ilp-equilibria}

A \cGAP\ program is \emph{linear} if all rules have the form:
\begin{eqnarray*}
A:\alpha_1X_1+\cdots+\alpha_nX_n & \leftarrow & B_1:X_1,\ldots,B_n:X_n.
\end{eqnarray*}
Linear GAPs were introduced in \cite{SNDOP_query} where the authors also showed that most classical diffusion models studied in the literature have linear approximations.

For any GAP $\Pi'$, Kifer and Subrahmanian \cite{ks92} show how to associate an operator $T_{\Pi'}$ that maps interpetations to interpretations as follows.
$T_{\Pi'}(I)(A) = \mathsf{sup}\{ \mu |\: A:\mu\leftarrow B_1:\mu_1,\ldots,B_n:\,u_n\}$ is a ground instance of a rule in $\Pi'$ and for all $1\leq i\leq n$,
$I(B_i)\geq\mu_i\}$. \cite{ks92} shows that $T_{\Pi'}$ is a monotonic operator on a complete lattice and
 that it has a least fixpoint $\mathsf{lfp}(T_{\Pi'})$. Moreover, $\mathsf{lfp}(T_{\Pi'})$ can be computed by iteratively applying the $T_{Pi'}$ operator.

Suppose $\bot$ is the interpetation that assigns 0 to all ground atoms. Then $T_{\Pi'}\uparrow 0 = \bot$. For successor ordinals $\alpha+1$, we set $T_{\Pi'}\uparrow (\alpha+1) = T_{\Pi'}(T_{\Pi'}\uparrow\alpha)$.
For limit ordinals $\alpha$, we set $T_{\Pi'}\uparrow\alpha(A) = \mathsf{sup}\{T_{\Pi'}\uparrow\beta(A)\: |\: \beta < \alpha\}$. By the Tarski-Knaster theorem, we are guaranteed that
$\mathsf{lfp}(T_{\Pi'}) = T_{\Pi'}\uparrow\alpha$ for some ordinal $\alpha$.
Additionally, Kifer and Subrahmanian~\cite{ks92} show that $\mathsf{lfp}(T_{\Pi'}) = {\cal MM}(\Pi')$.
A GAP is \emph{finitary} iff $\mathsf{lfp}(T_{\Pi'}) = T_{\Pi'}\uparrow k$ for some integer $k$.
We can extend this to \cGAP s as well.

\begin{definition}
A \cGAP\ $\Pi$ is finitary iff, for all interpretations $I$, the coherent transform of $\Pi$ w.r.t. $I$ (that is a GAP) is finitary.
\end{definition}

If we compute the minimal model of the GAPs $\Pi'$ in Step (9) of Algorithm~\ref{algo:allStable}, then we can develop an algorithm that uses mixed integer linear programs to compute the set of all strong equilibria of a \cGAP\ $\Pi$. The use of mixed integer linear programs for logic program computation was initially proposed by \cite{bell1994mixed,bell1996implementing} --- however, they did not consider either GAPs or Choice Logic Programs. 

Our Mixed Integer Linear Programming algorithm requires having an upper bound $\widehat{T}$ on the maximum number of iterations needed to compute the least fixpoint of $\Pi'$ for each possible choice in Algorithm~\ref{algo:allStable}.
We assume in the sequel that $\widehat{T}$ is such an upper bound --- later, in Section~\ref{sec:computeT}, we will provide a simple algorithm to compute $\widehat{T}$.
%\from{vs}{edoardo}{need to say somewhere in the paper how we get $T$.}

\subsection{Mixed Integer Linear Program (MILP)}
In this section we describe how to represent strong equilibria via a mixed integer linear formulation. 

\subsection{Variables in the MILP}
We start by defining the variables in the MILP we are constructing.

\begin{itemize}

\item $x^t_A$: For each $0 \leq t \leq \widehat{T}$ and
each ground atom $A$, the variable $x^t_{A} \in [0,1]$ denotes the value assigned to $A$ by $T_{\Pi'}\uparrow t$. Consequently, we fix 
$x^0_{A}=0$ for all $A$.

\item $z^t_i$: For each $1 \leq t \leq \widehat{T}$ and for each GAP rule $r_i: A_0:\mu_0 \leftarrow A_1:\mu_1,\dots,A_n:\mu_n$, the variable $z^t_i$ denotes the value assigned to 
$\mu_0$ by the rule $r_i$ in the interpretation  $T_{\Pi'}\uparrow (t-1)$.
\end{itemize}

The following example will be used in this section in order to help the reader understand the constraints used in our MILP.

\begin{example}\label{exspiega}
Consider the following ground cGAP program 

$$
\begin{array}{lcl}
r_1)\:\:a_1:0.3 & \leftarrow &\\
r_2)\:\:a_1:0.6 & \leftarrow &b_1:0.3\\
r_3)\:\:b_1,b_2 & \hookleftarrow & a_1,a_2
\end{array}$$ 

and assume that $\widehat{T}=2$. Then, we have the following real variables:
$$
\begin{array}{l}
x^0_{a_1},x^0_{a_2},x^0_{b_1},x^0_{b_2},
x^1_{a_1},x^1_{a_2},x^1_{b_1},x^1_{b_2},
x^2_{a_1},x^2_{a_2},x^2_{b_1},x^2_{b_2}\\
\\
z_1^1,z_2^1,
z_1^2,z_2^2\\
\end{array}
$$

Note that as $r_3$ is not a GAP rule, we do not consider it in the definition of the variables $z$.\hfill$\Box$
\end{example}

\subsection{Constraints in the MILP}
We now define the constraints of the MILP. There are 4 types of constraints in the MILP. We describe each type below and provide examples where appropriate.

\subsubsection{Type 1 Constraints}
For each atom $A$ and for each $1 \leq t \leq \widehat{T}$, let $rules(A) = \{r_{j_1},\dots,r_{j_h}\}$ 
be the set of all annotated rules in $\Pi'$ having $A$ in the head.

\begin{itemize}

\item If $rules(A) = \emptyset$ and $A$ does not appear in a head of a vertex choice rule, we add the constraint $x^t_A = 0$. This corresponds to the case where $A$ has no chance of being true as no ground rule has $A$ as its head.

\item If $rules(A) \neq \emptyset$
we model the fact that $x^t_A= \max_{r_i \in rules(A)} z^t_i$ by introducing a binary variable $u^t_i$ for each rule $r_i \in rules(A)$, and the following constraints:

\begin{eqnarray*}
z^t_i - x^t_{A} - u^t_i & \geq -1  & \forall i \in \{1,\dots,h\}\\
x^t_{A} - z^t_i & \geq 0 & \forall i \in \{1,\dots,h\}\\
\sum_{i=1}^h u^t_i & = 1 &
\end{eqnarray*}

In fact, as $x^t_{A} \geq z^t_i$ $\forall i \in \{1,\dots,h\}$ and there must exist only one rule $r_i$ for which $x^t_{A} \leq z^t_i$ (third and second constraints, respective), then $x^t_{A}$ will assume the value of the maximum $z^t_i$.
\end{itemize}

\begin{example}\label{exspiega1}
By considering the program in Example~\ref{exspiega}, we see that the only atom $A$ having $rules(A)\neq \emptyset$ is $a_1$ for which $rules(a_1) = \{r_1,r_2\}$. Moreover, only $a_2$ satisfies the condition for which $rules(a_2)=\emptyset$ and $a_2$ does not appear in the head of the vertex choice rule.
We therefore have the following set of constraints:
$$
\begin{array}{ccc}
x^t_{a_2}&=0&\forall t\in \{0,1,2\}\\
x^0_{a_1}&=0&\\
z^t_1- x^t_{a_1} - u^t_1 & \geq -1& \forall t\in \{1,2\}\\
z^t_2- x^t_{a_1} - u^t_2 & \geq -1& \forall t\in \{1,2\}\\
x^t_{a_1} - z^t_1 & \geq 0 &\forall t\in \{1,2\}\\
x^t_{a_1} - z^t_2 & \geq 0 &\forall t\in \{1,2\}\\
u^t_1 +u^t_2&=1&\\
u^t_1,u^t_2&\in \{0,1\}&
\end{array}
$$\hfill$\Box$
\end{example}

\subsubsection{Type 2 Constraints}
For each $1 \leq t \leq \widehat{T}$ and for each rule $r_i \in rules(A)$ of the form
$$r_i)\:\: A_0:f(\mu_1,\dots,\mu_n) \leftarrow A_1:\mu_1,\dots,A_{n}:\mu_{n},C_{1}:c_1,\dots,C_{l}:c_l$$
where $c_1,\dots,c_l$ are constants, and $C_{1}:c_1,\dots,C_{l}:c_l$ is a condition for the activation of the rule, 
we introduce 

\begin{itemize}
\item $l+1$ binary variables $v_1^t,\dots,v_l^t,v_{1,l}^t$, where each variable $v_j^t$, $1 \leq j \leq l$, checks if the condition 
$C_{j}:c_j$ is satisfied, and the binary variable $v_{1,l}$ checks if the whole condition $C_{1}:c_1,\dots,C_{l}:c_l$ is satisfied 

\item  the following constraints: 

\begin{eqnarray}
v^t_j - x^{t-1}_{C_j} & \geq \epsilon - c_j & \forall j \in \{1,\dots,l\}\\
x^{t-1}_{C_j} - v^{t}_j & \geq c_j - 1 & \forall j \in \{1,\dots,l\}\\
v^t_{j} - v^t_{1,l} & \geq 0 & \forall j \in \{1,\dots,l\}\\
v^t_{1,l} - \sum_{j=1}^{l} v^t_j & \geq 1 - l & \\
- z^t_i + v^t_{1,l} & \geq 0  &\\
z^t_i - f(x^{t-1}_{A_1},\dots,x^{t-1}_{A_n}) - v^t_{1,l} & \geq -1  & \\
f(x^{t-1}_{A_1},\dots,x^{t-1}_{A_n}) - z^t_i - v^t_{1,l} & \geq -1  & 
\end{eqnarray}

where $\epsilon$ is a very small real number used to map the strictly greater constraints.
Constraint (1) imposes that  $v^t_j=1$ if $x^{t-1}_{C_j} \geq c_j$, while Constraint (2) imposes that $v^t_j=0$ if $x^{t-1}_{C_j} < c_j$.
Observe that the above constraints are correct for $c_j^t \geq \epsilon$. If $c_j^t = 0$,
then $c_j^t$ can be removed from the constraints.
Constraint (3) requires that if there exists a $v^t_j=0$, i.e. the whole condition is not satisfies, then $v^t_{1,l}=0$, while Constraint (4)
imposes that if $v^t_j=1$, for $1 \leq j \leq l$, i.e. the whole condition is satisfied, then $v^t_{1,l}=1$.
Constraint (5) imposes that, if  $v^t_{1,l}=0$, then $z^t_i=0$, while if $v^t_{1,l}=1$, from Constraints (6) and (7) we have that 
$z^t_i=f(x^{t-1}_{A_1},\dots,x^{t-1}_{A_n})$.

Observe that the values of variables $z^t_i$ computed at step $t$ depend on the values of the variables  $x^{t-1}_{A_1},\dots,x^{t-1}_{A_n}$
computed at step $t-1$.
\end{itemize}
We now present an example of Type (2) constraints.

\begin{example}
Consider the rule $$r_2)\:\:a_1(1):0.6  \leftarrow b_1(1):0.3$$ 
of the program in Example~\ref{exspiega}. We have just one condition $b_1(1):0.3$. Hence,
we have the following set of constraints:
$$
\begin{array}{ccc}
v^t_1 - x^{t-1}_{b_1}&  \geq \epsilon - 0.3&  \forall t \in \{1,2\}\\
x^{t-1}_{b_1} - v^{t}_1 & \geq 0,3 - 1 & \forall t \in \{1,2\}\\
v^t_{1} - v^t_{1,1} & \geq 0 & \forall t \in \{1,2\}\\
v^t_{1,1} - v^t_1 & \geq 0 & \forall t \in \{1,2\}\\
- z^t_2 + v^t_{1,1} & \geq 0  &\forall t \in \{1,2\}\\
z^t_2 - 0.6 - v^t_{1,1} & \geq -1  & \forall  t \in \{1,2\}\\
0.6 - z^t_2 - v^t_{1,1} & \geq -1  & \forall t \in \{1,2\} \\
v^t_{1}, v^t_{1,1}&\in \{0,1\}&
\end{array}
$$\hfill$\Box$
\end{example}

\subsubsection{Type 3 Constraints} 
For each ground vertex choice rule $r_p$ of the form $r_p: B_1,\dots,B_m \hookleftarrow A_1,\dots,A_m$,
we introduce $m$ binary variables $y_{r_p}^1,\dots,y_{r_p}^m$, and the following constraints:
\begin{eqnarray*}
\sum_{i=1}^m y_{r_p}^i & = 1 &
\end{eqnarray*}
and for each $1 \leq t \leq T$
\begin{eqnarray*}
x^t_{B_i} - x^{t}_{A_i} - y_{r_p}^i & \geq -1  & \forall i \in \{1,\dots,m\}\\
x^{t}_{A_i} - x^{t}_{B_i} - y_{r_p}^i & \geq -1  & \forall i \in \{1,\dots,m\}\\
x^t_{B_i} + y_{r_p}^i & \geq 0  & \forall i \in \{1,\dots,m\}\\
- x^t_{B_i} + y_{r_p}^i & \geq 0  & \forall i \in \{1,\dots,m\}\\
%(\ \sum_{i=1}^m x^t_{B_i} & \leq 1 &\\
%x^t_{B_i} & \leq x^{t-1}_{A_i}\ )&
\end{eqnarray*}

The above set of constraints means that only one variable $y_{r_p}^i$ can be $1$ (and the others must be 0), and
if $ y_{r_p}^i=1$ then $x^t_{B_i} = x^{t}_{A_i}$, otherwise $x^t_{B_i} = 0$.

\begin{example}
By considering the VC-rule $$r_3)\:\:b_1,b_2 \hookleftarrow  a_1,a_2$$ 
of the program in Example~\ref{exspiega},  we have the following set of constraints:
$$
\begin{array}{ccc}
y_{r_3}^1+y_{r_3}^2 & = 1 &\\
x^t_{b_1} - x^{t}_{a_1} - y_{r_3}^1 & \geq -1  &\forall t \in \{1,2\} \\
x^t_{b_2} - x^{t}_{a_2} - y_{r_3}^2& \geq -1  & \forall t \in \{1,2\} \\
x^{t}_{a_1} - x^{t}_{b_1} - y_{r_3}^1 & \geq -1  & \forall t \in \{1,2\} \\
x^{t}_{a_2} - x^{t}_{b_2} - y_{r_3}^2 & \geq -1  & \forall t \in \{1,2\} \\
x^t_{b_1} + y_{r_3}^1 & \geq 0  & \forall t \in \{1,2\} \\
x^t_{b_2} + y_{r_3}^1 & \geq 0  & \forall t \in \{1,2\} \\
- x^t_{b_1} + y_{r_3}^1 & \geq 0  & \forall t \in \{1,2\} \\
- x^t_{b_2} + y_{r_3}^2 & \geq 0  & \forall t \in \{1,2\} \\
y_{r_3}^1,y_{r_3}^2&\in \{0,1\}&\\
\end{array}
$$\hfill$\Box$
\end{example}

\subsubsection{Type 4 Constraints}

All the previous constraints guarantee that each feasible solution for them is a coherent model for the \cGAP.
To guarantee that such a coherent model is also a \SE\ we have to add the following constraints for each vertex choice rule
$r_c$ of the form $r_c: B_1,\dots,B_m \hookleftarrow A_1,\dots,A_m$:
\begin{eqnarray*}
- x^{\widehat{T}}_{A_j} + \sum_{i=1}^m x^{\widehat{T}}_{B_i} & \geq 0 & \forall j \in \{1,\dots,m\}
\end{eqnarray*}

Since only one $x^{\widehat{T}}_{B_j}$, $1 \leq j \leq m$, can be greater than or equal to $0$, and the others must be $0$,  then $\sum_{i=1}^m x^{\widehat{T}}_{B_i} = x^{\widehat{T}}_{B_j}$ and the constraints impose that $x^{\widehat{T}}_{B_i} \geq  x^{\widehat{T}}_{A_j}$, $\forall j \in \{1,\dots,m\}$, i.e. the condition of \SE\ is satisfied.
\begin{example}\label{exspiega5}
By considering the program in Example~\ref{exspiega}, we have that the conditions establishing the \SE\ are:
$$
\begin{array}{ccc}
- x^{2}_{a_1} + x^{2}_{b_1} +x^{2}_{b_2} & \geq 0 & \\
- x^{2}_{a_2} + x^{2}_{b_1} +x^{2}_{b_2} & \geq 0 & \\
\end{array}
$$\hfill$\Box$
\end{example}

\begin{algorithm}[t]
\caption{Algorithm computing $\widehat{T}$.}\label{computeT}
\begin{algorithmic}[1]
\Procedure{$compute\widehat{T}$}{\cGAP\ program $\Pi$, $\delta$}
\State $\overline{T}=\delta$
\While{($oracle(\Pi,\overline{T})=false$)} 
\State $\overline{T} = \overline{T} + \delta$
\EndWhile
\State $min=\overline{T}-\delta$, $max=\overline{T}$
\State $t =  \lfloor(min+max)/2\rfloor$
\While{($min < max -1$)}
\If{($oracle(\Pi,t)=true$)} 
\State $max=t$
\Else \ $min = t$
\EndIf
\State $t =  \lfloor(min+max)/2\rfloor$
\EndWhile
\State \Return $max$
\EndProcedure
\Statex
\Procedure{$oracle$}{\cGAP\ program $\Pi$, $\overline{T}$}
\State Solve $\max \sum_{A \ ground \ atom \ in \ \Pi}  x_A^{\overline{T}} - x_A^{\overline{T}-1}$  s.t. $ILC(\Pi,\overline{T})$
\If{($\sum_{(A \ ground \ atom \ in \ \Pi)}  x_A^{\overline{T}} - x_A^{\overline{T}-1} = 0$)} 
\State \Return $true$
\EndIf
\State \Return $false$
\EndProcedure
\end{algorithmic}
\end{algorithm}

Given a \cGAP\ $\Pi$ and the maximum number of iterations $\widehat{T}$, let $ILC(\Pi,\widehat{T})$ denote the set of constraints listed above, i.e. the constraints of types (1) -- (4).
From the constraint description provided above, we can derive the following trivial propositions.

\begin{rproposition}
Let $\Pi$ be a finitary linear GAP.
The set of type (1) and (2) constraints has a unique solution that coincides with the least fixpoint model of $\Pi$.~\hfill~$\Box$
\end{rproposition}

\begin{rproposition}
Let $\Pi$ be a finitary linear \cGAP.
Every solution for the set of
Type (1), (2), and (3) constraints represents a coherent model for $\Pi$.~\hfill~$\Box$
\end{rproposition}

\begin{rproposition}~\label{prop:ILCallSE}
Let $\Pi$ be a finitary linear \cGAP.
Every solution for the set $ILC(\Pi,\widehat{T})$ of constraints represents a \SE\ of $\Pi$.~\hfill~$\Box$
\end{rproposition}
The above theorems capture the fact that the constraints in $ILC(\Pi,\widehat{T})$ neatly capture  strong equilibria of $\Pi$.

\subsection{Algorithm to find $\widehat{T}$: the maximum number of iterations needed} \label{sec:computeT}  
Algorithm~\ref{computeT} provides a method to compute the maximum number of iterations $\widehat{T}$ referenced above.
Basically, Algorithm~\ref{computeT} uses a oracle that checks whether all the solutions of constraints $ILC(\Pi,\widehat{T})$ represent fixpoints, and 
finds the minimum value $\widehat{T}$ for which the oracle answer is true. Formally, the oracle returns true if $\sum_{(A \ ground \ atom \ in \ \Pi)}  x_A^{\widehat{T}} - x_A^{\widehat{T}-1} = 0$, i.e. does not exist a solution for which the fixpoint is not reached, while it returns false if 
$\sum_{(A \ ground \ atom \ in \ \Pi)}  x_A^{\widehat{T}} - x_A^{\widehat{T}-1} > 0$, i.e. there exists a solution for which the fixpoint is not reached. Then, Procedure $compute\widehat{T}$ increases $\overline{T}$ by a quantity $\delta$ given in input, until the oracle returns ``true'' as the answer. Then, by using a binary search, it finds the minimum value of $t \in (\overline{T}-\delta,\overline{T}]$ for which the oracle still answers true.

%\textbf{NOTE}: If the program is VIC we can do further optimizations over the ILP resolution algorithm.

\subsection{Algorithm to find all \SEs}
We now present an algorithm (Algorithm~\ref{enumerate}) to find all strong equilibria of a \cGAP\ $\Pi$ by using $ILC(\Pi)$.

\begin{algorithm}[t]
\caption{Algorithm enumerating all \SEs.}\label{enumerate}
\begin{algorithmic}[1]
\Procedure{$enumerate$}{\cGAP\ program $\Pi$}
\State Let $\Sigma = ILC(\Pi)$ and $Sol = \{\}$.
\While{A feasible solution $s$ for $\Sigma$ exists}
\State $Sol = Sol \cup \{s\}$.
\State Let $k^1_{r_p},\dots,k^m_{r_p}$ the values assumed in $s$ by the variables $y^1_{r_p},\dots,y^m_{r_p}$ %associated with the player $p$.
\State Add to $\Sigma$ the constraint\label{newC}
$$\sum_{i=1}^{|\Gamma_{\Pi}|}\sum_{j=1}^m y_{r_i}^j(1 - k_{r_i}^j) + \sum_{i=1}^{|\Gamma_{\Pi}|} \sum_{j=1}^m (1 - y_{r_i}^j)k_{r_i}^j \geq 1$$
\EndWhile\\
\Return $Sol$
\EndProcedure
\end{algorithmic}
\end{algorithm}

According to Proposition~\ref{prop:ILCallSE}, strong equilibria of $\Pi$ correspond exactly to solutions of $ILC(\Pi,\widehat{T})$. Algorithm~\ref{enumerate} enumerates the set of all strong equilibria by first finding one solution of $ILC(\Pi,\widehat{T})$ (step 3).
It then adds a constraint (step 6) that performs a ``cut'' on the convex polytope defined by $ILC(\Pi,\widehat{T})$. This cut eliminates the strong equilibrium $s$ and no other strong equilibria. We then add this constraint to $ILC(\Pi,\widehat{T})$
and continue the process of solving the now expanded $ILC(\Pi,\widehat{T})$ till $ILC(\Pi,\widehat{T})$ becomes unsolvable.

%An alternative algorithm can be the one of enumerate all solutions by means of the branch and bound (or cut) tree used by the ILP solver. It easy to see that such algorithm requires polynomial space.

\section{Vertex Independent Choice Programs}\label{sec:vic-programs}

As strong equilibria may not exist for all \cGAPs\, we will define a class of programs called 
 \emph{vertex independent choice (VIC) programs},
for which a \SE\ always exists when the size of the vertex choice rule is $2$. We use $VIC_2$ to denote this class of programs.

\begin{definition}[Dependency graph]
Suppose $\Pi$ is a \cGAP. 
The dependency graph $\G(\Pi)$ associated with $\Pi$ has the set $VP$ of vertex predicates
as the set of vertices. The set $E$ of edges is defined as follows:
 $(p_2,p_1) \in E$ iff
\begin{itemize}
\item there is a \cGAP \ rule $r$
with $p_2$ appearing in $body(r)$ and $p_1$ in $head(r)$, or
\item in the vertex choice rule $$r: B_1,\dots,B_m
\hookleftarrow A_1,\dots,A_m$$
there is an $1\leq i\leq m$ such that $p_2$ appears in $A_i$ and $p_1$ appears in $B_i$.\hfill$\Box$
\end{itemize}
\end{definition}
We are now ready to define a VIC program.

\begin{definition}[Vertex Independent Choice (VIC) program]
A \cGAP \  $\Pi$ is said to be \emph{Vertex Independent Choice (VIC)} if
\begin{enumerate}
  \item every predicate symbol appearing in the head of the VC-rule in $\Pi$ does not
appear in the head of a GAP rule, and
  \item Suppose $B_1,\dots,B_m \hookleftarrow A_1,\dots,A_m \in \Pi$ and $p_1$
appears in $B_j$ and $p_2$ appears in $A_i$ and $i \neq j$. Then there is no path from 
 $p_1$ to $p_2$ in the dependency graph
$\G(\Pi)$.\hfill$\Box$
\end{enumerate}
A VIC-program is said to be a $VIC_m$ program when its VC-rule has the form $B_1,\dots,B_m \hookleftarrow A_1,\dots,A_m$.
\end{definition}

Intuitively, the VIC condition requires that the choice of a vertex is completely
independent, because (1) it cannot be forced by factors other than 
the diffusion process, and (2) it is not influenced by conflicting atoms.
Note that the Labour-Tory example in Section~\ref{sec:example} is a VIC$_2$ program.

Given a VIC$_m$ program $\Pi$ containing the vertex choice rule $b_1(X),\dots,b_m(X)$ $\hookleftarrow a_1(X),\dots,a_m(X)$,  and having dependency graph $\G(\Pi)$, we  define $m$ sets of predicates
${Pred}_\Pi^1,\dots,{Pred}_\Pi^m$ of $\Pi$, such that each set ${Pred}_\Pi^i$ contains all the predicates obtained by a reverse visit (i.e with each edge inverted, e.g. edge $(a,b)$ is considered as $(b,a)$) of $\G(\Pi)$ starting from the predicate $b_i$.
% Analogously, from the set $\atoms$ of ground atoms of $\Pi$,
% we can always generate $m$ sets of atoms ${At}_\Pi^1,\dots,{At}_\Pi^m$ of $\atoms$, where each ${At}_\Pi^i$ contains all ground atoms using predicates in ${Pred}_\Pi^i$.
Moreover, given a state $S$, we can divide the induced ground VIC program $\Pi_{S}$ into $m$
independent programs $\Pi^1_{S},\dots,\Pi^m_{S}$, where each $\Pi^m_{S}$ contains all rules from $\Pi_{S}$ involving only predicates from ${Pred}^i_{\Pi_{S}}$.

% Given couple $(p_r(v),i)$, where $p_r(v)$ is a player and $i$ is an action in $Q(p_r(v))$,
% we define the \emph{dependency set of $p_r(v)$ and $i$}, denoted $d((p_r(v),i))$,
% as the set of the couples $(p_{r'}(v'),j)$ such that there exists a path from
% $B_i$ and $B'_j$ in the dependency graph,
% where $r: B_1,\dots,B_m \hookleftarrow A_1,\dots,A_m$ and
% $r': B'_1,\dots,B'_{m'} \hookleftarrow A'_1,\dots,A'_{m'}$.
% We also include the pair $(p_r(v),i)$ in $d((p_r(v),i))$.
%
% \begin{proposition}
% Given a VIC program P, each set $d(p,j)$ contains, for each player $p' \in \Gamma_{\Pi}$, at most one pair $(p',i)$, where $i \in Q(p')$
% \end{proposition}
%
% \begin{proposition}\label{prop:dep}
% Given a VIC program P and a player $p$, $d(p,i) \cap d(p,j) = \emptyset$ if $i \neq j$.
% \end{proposition}

\begin{example}
Let $\Pi$ be the VIC$_2$ program shown in Section~\ref{sec:example}. We have two sets of predicates, i.e.
$$\begin{array}{cl}
{Pred}_\Pi^{L}=\{&voteLabour^U,voteLabour^D,suptBrown,knows,\\ &mentor,student,olderRel\}\\
{Pred}_\Pi^{T}=\{&voteTory^U,voteTory^U,likeCam,mentor,\\&employee,idol,young\}\\
\end{array}$$
Since the vertex choice rule is 
$$voteTory^{D}(X),voteLabour^{D}(X)  \hookleftarrow 
voteTory^U(X),voteLabour^U(X)$$
we have that $Q=\{1,2\}$, where $1$ is associated with the choice $voteTory$, while $2$ is associated with the choice $voteLabour$. 
Consider the state $S$ where the choice for player $A$  is $S(A)=1$ ($voteTory$), while the choice
for player $B$  is $S(B)=2$ ($voteLabour$).
Then the programs $\Pi^L_S$ and $\Pi^T_S$ are
$$
\begin{array}{lrcl}
\Pi^L_S:&voteLabour^U(A):X & \leftarrow & suptBrown(A):X\\
&voteLabour^U(B):X & \leftarrow & voteLabour^{D}(A):X, \ knows(B,A):1\\
&voteLabour^U(B):X & \leftarrow & voteLabour^{D}(A):X, \ mentor(B,A):1,\\
&& & student(B):1\\
&voteLabour^U(B):X & \leftarrow & voteLabour^{D}(A):X, \ olderRel(B,A):1\\
&voteLabour^{D}(B) & \leftarrow &
voteLabour^U(B)\\
&&&\\
\Pi^T_S:&voteTory^U(A):X & \leftarrow & likeCam(A):X\\
&voteTory^U(B):X & \leftarrow & voteTory^{D}(A):X, \ mentor(B,A):1,\\
&& & employee(B):1\\
&voteTory^U(B):X & \leftarrow & voteTory^{D}(A):X, \ idol(B,A):1,\\
& && young(B):1\\
&voteTory^{D}(A) & \leftarrow &
voteTory^U(A)\\
\end{array}
$$\hfill$\Box$
\end{example}

The following result shows some properties of VIC programs.

\begin{rproposition}[\proputility]\label{prop:utility}
Given two states $S_1$ and $S_2$ of a VIC program $\Pi$,
where $S_2$ only differs from $S_1$ in the choice of player $p$, i.e.
$S_1(p') = S_2(p')$ if $p' \neq p$, and $S_1(p') \neq S_2(p')$ if $p' = p$,
then for each player $\hat{p}\in \Gamma_{\Pi}$ the following statements hold:
\begin{itemize}
 \item[1.] $u(S_1,\hat{p},S_1(p))\geq  u(S_2,\hat{p},S_1(p))$.
 \item[2.] $u(S_1,\hat{p},S_2(p))\leq  u(S_2,\hat{p},S_2(p))$.
 \item[3.] $\forall j\in Q\setminus \{S_1(p),S_2(p)\}: \ \ u(S_1,\hat{p},j)=u(S_2,\hat{p},j)$.~\hfill~$\Box$
\end{itemize}
\end{rproposition}

Example~\ref{exa:vice} below shows that VIC programs are not guaranteed to have strong equilibria.

\begin{example}\label{exa:vice}
Consider the following VIC program where the size of vertex choice rule is $3$.
\[
\begin{array}{rcl}
  g^U(1):0.4 & \leftarrow &  \\
  r^U(2):0.4 & \leftarrow &  \\
  b^U(3):0.4 & \leftarrow &  \\\\
  b^U(1):1.0 & \leftarrow & b^D(3):0.2\\
  g^U(2):1.0 & \leftarrow & g^D(1):0.2\\
  r^U(3):1.0 & \leftarrow & r^D(2):0.2\\\\
  g^D(X),r^D(X),b^D(X) & \hookleftarrow & g^U(X),r^U(X),b^U(X)
\end{array}
\]
This program does not have any \SE. Moreover, observe that if we remove any one of the facts, three \SEs\ exist.
\end{example}

The following result shows that the problem of checking existence of a strong equilibrium for $VIC_3$ programs is NP-hard.

\begin{rtheorem}[\thconsistCheck][Existence of Strong Equilibrium for $VIC_3$ programs]\label{th:consistCheck}
\ \\ Given a VIC \cGAP \ program $\Pi$ where the size of vertex choice rule is $3$, the problem of deciding whether $\Pi$ has a \SE\ is still {\bf NP}-hard under data and combined complexity.~\hfill~$\Box$
\end{rtheorem}

%confronto VIC STABLE EQUILIBRIUM MODEL contro VIC NASH EQUILIBRIUM
%VIC STABLE EQUILIBRIUM MODELs contengono VIC NASH EQUILIBRIUM MODELs
%VIC NASH EQUILIBRIUM MODELs esistono sempre ?

The following result shows that for VIC programs, all Nash equilibria are strong equilibria, but the converse is not necessarily true.

\begin{rtheorem}[\thmEverySEisNE]
Let $\Pi$ be a VIC program. Then every Nash equilibrium is a \SE\ for $\Pi$, but in general a \SE\ for $\Pi$ may not be a Nash equilibrium.~\hfill~$\Box$
\end{rtheorem}

\subsection{VIC$_2$ Programs}
Fortunately, $VIC_2$ programs have two nice properties. First, they are guaranteed to have a strong equilibrium. And second, the problem of finding a strong equilibrium can be solved in polynomial time.

Algorithm~\ref{SEalgo} shows how to find such a strong equilibrium. We use the concept of state defined in Section~\ref{sec:game-formalization}. We start (line $2$) by creating an initial state where all players take action $1$
(of the two actions $1,2$ supported by the $VIC_2$ program). Recall that for each player $\calpv$, we have only two choices in $VIC_2$ programs, i.e. $Q(\calpv) = \{1,2\}$. If this state is not a strong equilibrium, we identify all players for which a higher utility is obtained by performing action 2 (lines $3-7$) and if this is the case, we set their action appropriately. Finally (line $8$), we return the minimal model of the induced ground GAP $\Pi_S$ (see Definition~\ref{def:inducedGAP}).
%\from{vs}{all}{algorithm needs some comments in it to explain notation which is very confusing.}

\begin{algorithm}[t]
\caption{Algorithm finding a \SE.}\label{SEalgo}
\begin{algorithmic}[1]
\Procedure{$findSE$}{\cGAP\ $VIC_2$ program $\Pi$}
\State Let $S$ be a state s.t. $S(\calpv)=1$ for all players $\calpv \in \Gamma_{\Pi}$;
%\ForAll{($\calpv \in \Gamma_{\Pi}$)}
%\State Set $S(\calpv)=1;$
%\EndFor
\While{($S$ does not represent a \SE\ for $\Pi$)}
\ForAll{(players $\calpv$ s.t. $u(S,\calpv,1) < u(S,\calpv,2)$)}
\State Set $S(\calpv)=2;$
\EndFor
\EndWhile \\
\Return ${\cal MM}(\Pi_S)$
\EndProcedure
\end{algorithmic}
\end{algorithm}

Observe that a different \SE\ can be found by inverting
action $1$ with $2$ and vice versa.

\begin{rtheorem}[\thmAlgoFour]\label{th:8.6}
Algorithm~\ref{SEalgo} runs in ${\bf PTIME}$ and returns a \SE\ (that always exists).~\hfill~$\Box$
\end{rtheorem}

From the set of all ground atoms of a $VIC_2$ program $\Pi$, we can define two sets: ${\cal MM}(\Pi^1_S)$ is the set of all actions in
${\cal MM}(\Pi_S)$  atoms linked with the action $1$ --- ${\cal MM}(\Pi^2_S)$ is the set of all actions in
${\cal MM}(\Pi_S)$  atoms linked with the action.
%For each state $S$ we call this two set respectively, ${\cal MM}(\Pi^1_S)$ and ${\cal MM}(\Pi^2_S)$.
Let $S_{12}$ ($S_{21}$) be the state identifying the \SE\ computed by Algorithm~\ref{SEalgo} (resp. by inverting the action $1$ with $2$ and vice versa).
The following result shows certain relationships about the utilities returned by the different minimal models of GAPs depending upon our choice of $S$.

\begin{rtheorem}[\thmMaxMinModels][Maximal and Minimal models]\label{maxmin}
 Suppose $\Pi$ is a $VIC_2$ program. For each state $S$ identifying a \SE\ the following statements hold:
 \begin{itemize}
  \item[1.] ${\cal MM}(\Pi^2_{S_{12}}) \preceq {\cal MM}(\Pi^2_S) \preceq {\cal MM}(\Pi^2_{S_{21}})$
  \item[2.] ${\cal MM}(\Pi^1_{S_{21}}) \preceq {\cal MM}(\Pi^1_S) \preceq {\cal MM}(\Pi^1_{S_{12}})$~\hfill~$\Box$
 \end{itemize}
\end{rtheorem}

The following result shows that checking whether an action is true in all strong equilibria is polynomially solvable in the case of $VIC_2$ programs.

\begin{rproposition}[\propEntailVICTwo][Entailment in $VIC_2$ ]
Given a $VIC_2$ \ program $\Pi$ and a ground annotated atom $AA$, the problem of deciding whether $\Pi \models AA$ is in ${\bf PTIME}$ under data and combined complexity.~\hfill~$\Box$ %(i.e in the size of the embedded Social Network).
% under combined complexity is also complete.
\end{rproposition}

The result below, however, states that checking for existence of strong equilibria different from ${\cal MM}(\Pi_{S_{21}})$ and ${\cal MM}(\Pi_{S_{12}})$ is still computationally intractable.

\begin{rtheorem}[\thmVICtwoHard]\label{th:vic2hard}
The problem of deciding whether there exists a \SE \ $M$ different from ${\cal MM}(\Pi_{S_{21}})$ and ${\cal MM}(\Pi_{S_{12}})$ is {\bf NP}-hard under data and combined complexity.~\hfill~$\Box$
\end{rtheorem}

The following result shows complexity results for counting problems associated with $VIC_2$ programs.

\begin{rtheorem}[\thmCountingComplexityVIC]\label{th:countingComplexityVIC2}\
\begin{itemize}
\item Given a VIC$_2$ program $\Pi$, the problem of counting the number of \SEs\ is {\bf \#P}-hard under data and combined complexity.
\item Given a positive integer $k$ and VIC$_2$ program $\Pi$, the problem of deciding whether $\Pi$ has at least $k$ \SEs\ is {\bf PP}-hard.~\hfill~$\Box$
\end{itemize}
\end{rtheorem}

At this point, since the number of \SEs\ for a VIC$_2$ program can be exponential, the question is whether it is possible to enumerate all the \SEs\ by using a polynomial total time algorithm. We recall that
an algorithm generating all configurations that satisfy a given specification is said to be \emph{polynomial total time} \cite{PapaIPL88} if the time required to output all configurations is bounded by a polynomial in $n$ (the size of the input) and $C$ (the number of configurations). Unfortunately, as stated in Proposition \ref{prop:noPTT}, this is not possible unless ${\bf P} = {\bf NP}$.

\begin{rproposition}[\propNoPTT]\label{prop:noPTT}
If there exists a polynomial-time algorithm for generating all the \SEs\ of a VIC$_2$ program, then ${\bf P} = {\bf NP}$.~\hfill~$\Box$
\end{rproposition}

\section{Estimation Queries and Competitive Diffusion}\label{sec:comp-diff-est}
In this section, we look at the problem of estimating the diffusion of concepts when competing diffusions are involved. We view these as queries that we call ``estimation queries''.

Let $\Pi$ be a \cGAP\ and $b_1(X),\dots,b_m(X) \hookleftarrow a_1(X),\dots,a_m(X)$ be its vertex choice rule. Then, for each atom $b_i(X)$, $i=1,\dots,m$, we have an additional \emph{choice} atom $c\_b_i(X)$, assuming a
value in the (binary) set $\{0,1\}$, specifying whether the player $X$
chose the action $b_i$. Given an interpretation $I$, let $I(c\_b_i(X)) = 1$ if player $X$ chose action $b_i$ --- otherwise $I(c\_b_i(X)) = 0$.
We use $C(\Pi)$ to denote the set of all choice atoms $c\_b_i(X)$.
Choice atoms have been introduced in order to correctly compute queries involving estimated diffusion through the network.

\begin{definition}[Estimation Query] An estimation query (or just query for short) $\cal{Q}$ over a \cGAP\ $\Pi$ is a pair
$${\cal Q} = \langle f(\mu_1,\dots,\mu_n), \{A_1:\mu_1,\dots,A_n:\mu_n\} \rangle$$
where $f(\mu_1,\dots,\mu_n)$ is a generic function computable in PTIME, and $\{A_1:\mu_1,\dots,$ $A_n:\mu_n\}$ is a collection of ground atoms of $\Pi \cup C(\Pi)$.
\end{definition}
Intuitively, this estimation query takes as input, a set $\{ A_1:\mu_1,\ldots,A_n:\mu_n\}$ of ground annotated atoms and a function that aggregates the $\mu_i$'s together. The annotated atoms specify which annotated atoms are important in estimating some quantity, and the function $f$ specifies how these annotations are to be merged together into a single annotation.
The function $f$ is often an aggregation function such as {\tt sum}, {\tt count}, etc. The following example provides an illustration of this concept.

\begin{example}\label{ex:count}
Consider the example about the Labour-Tory election presented in Section~\ref{sec:example}, and suppose we are interested in estimating
the number of persons voting for Labour. This can be expressed as the estimation estimation query
$${\cal Q} = \langle {\tt sum}(\mu), \{c\_voteLabour^D(X):\mu \ | \ X \in \V \} \rangle$$
We use the function ${\tt sum}$ instead of ${\tt count}$ since the value $\mu$ assumed by each of the atoms $c\_voteLabour^D(X)$ can be only $0$ or $1$.
\hfill$\Box$
\end{example}

Let $\Pi$ be a \cGAP, and let ${\cal Q}$ be an estimation query over $\Pi$.
We say that a value $a$ is a \emph{possible answer} to estimation query $\cal{Q}$ w.r.t. a \cGAP\ $\Pi$
if there is a \SE\ $M$ for $\Pi$ such that $f(M(A_1),\dots,M(A_n)) = a$.
We write $poss(\Pi,\cal{Q})$ to denote the set of all possible
answers of the estimation query $\cal{Q}$ over $\Pi$.

We consider the following semantics to the estimation query $\cal{Q}$ over $\Pi$.

\begin{definition}[Range Answer]
The \emph{range answer} is the interval $$[L_
{\cal{Q}}; U_{\cal{Q}}] = [{\tt glb}(poss(\Pi,{\cal{Q}}));{\tt lub}(poss(\Pi,{\cal{Q}}))]$$
where {\tt glb} and {\tt lub} stand, respectively, for greatest lower bound and
least upper bound.
\end{definition}

\begin{example}\label{ex:17}
Consider the VIC$_2$ cGAP $\Pi$ of Example~\ref{ex:2eq}, where we have only two \SEs, namely $SE_a$ and $SE_b$:

\begin{center} \scriptsize
\begin{tabular}{|c|c|c|c|c|}
\hline
&$buyAsus^U(1)$ & $buyAsus^D(1)$ & $buyMac^U(1)$ & $buyMac^D(1)$  \\ \hline
$SE_a$ &$0.6$ &$0.6$ & $0.3$ & $0.0$\\
$SE_b$ &$0.3$ & $0.0$ & $0.3$ & $0.3$ \\ 
\hline
\end{tabular}
\end{center}

Suppose we want to know the confidence that vertex $1$ has in buying an $Asus$, i.e. our estimation estimation query is 
${\cal Q} = \langle {\tt sum}(\mu), \{buyAsus^D(1):\mu\} \rangle$. We have that $poss(\Pi,{\cal Q})=\{0.3,0.6\}$, and then
${\tt glb}(poss(\Pi,{\cal{Q}}))=0.3$ and  ${\tt lub}(poss(\Pi,{\cal{Q}}))=0.6$. Thus, the range answer to the estimation query ${\cal Q}$ is the 
interval $[0.3,0.6]$.

By considering now the estimation query ${\cal Q'} = \langle 1- {\tt sum}(\mu), \{buyAsus^D(1):\mu\} \rangle$ we have that the range answer is the interval $[0.4,0.7]$.\hfill$\Box$
\end{example}

The following result says that the complexity of determining the range answer to an estimation query is intractable.

\begin{rtheorem}[\thmQueryOne]\label{th:estimationQueryComplexity}
Given a \cGAP\ $\Pi$, an estimation query ${\cal{Q}}$ over $\Pi$ and a possible answer $a$ of ${\cal{Q}}$, the problem of decide whether $a = {\tt glb}(poss(\Pi,{\cal{Q}}))$
(resp. $a = {\tt lub}(poss(\Pi,{\cal{Q}}))$) is in {\bf DP} and co-{\bf NP}-hard.~\hfill~$\Box$
\end{rtheorem}

%\begin{theorem}
%Given a \cGAP\ $\Pi$, and an estimation query $Q$ over $\Pi$, the problem of finding ${\tt glb}(poss(\Pi,Q))$ (resp. ${\tt lub}(poss(\Pi,Q))$) is in ${\bf FP^{NP}}$.
%\end{theorem}

\begin{algorithm}[t]
\caption{Naive range answer computation.}\label{alg:range-ans}
\begin{algorithmic}[1]
\Procedure{$rangeAns$}{\cGAP\ $\Pi$, ${\cal{Q}} = \langle f(\mu_1,\dots,\mu_n), \{A_1:\mu_1,\dots,A_n:\mu_n\} \rangle$}
\State Compute the set $S_M$ of all coherent models by using Algorithm \ref{algo:allStable}
\State $glb=-\infty$ $lub=+\infty$
\For{each $M \in S_M$ }
\If{M is a strong equilibrium} 
\State Let $q$ be the result of computing ${\cal{Q}}$ over $M$
\If{$q > lub$} $lub=q$
\EndIf
\If{$q < glb$} $glb=q$
\EndIf
\EndIf
\EndFor
\State \Return $[glb,lub]$
\EndProcedure
\end{algorithmic}
\end{algorithm}

Algorithm~\ref{alg:range-ans} shows a straightforward method to compute the range answer to an estimation query.
However, by constructing an appropriate objective function for the MILP defined by constraints $ILC(\Pi,\widehat{T})$ from Section~\ref{sec:ilp-equilibria}, it is possible to solve the entailment problem and compute the range answer for estimation queries where the function $f$ is ${\tt sum}$, ${\tt count}$, ${\tt min}$, or ${\tt max}$. Even though ${\tt count}$, ${\tt min}$ and ${\tt max}$ are not linear functions, the expressive power of our  MILPs allow us to express them by adding other binary variables and constraints. In fact, using Type~1 constraints, where we have the binary variables $u^t_i$, for each rule $r_i \in rules(A)$, we are able to model  the nonlinear constraint $x^t_A= \max_{r_i \in rules(A)} z^t_i$.
This means that we can  also handle \cGAPs\ whose annotation functions are nonlinear.

\subsection{Monotone  estimation queries}
In this section, we show an important result. The range answer to queries that are monotone (which include most things we would probably want to know with respect to diffusion in social networks including ${\tt sum}$, ${\tt count}$, ${\tt min}$, or ${\tt max}$) can be computed in polynomial time.

\begin{definition}[Monotone estimation query]
An estimation query ${\cal{Q}} = \langle f(\mu_1,\dots,$ $\mu_n), A_1:\mu_1,\dots,A_n:\mu_n \rangle$ over a linear \cGAP\ $\Pi$ is \emph{monotone} if, given two real vectors $\bar{x} = [x_1,\dots,x_n]$ and $\bar{y} = [y_1,\dots,y_n]$ s.t.
$x_i \leq y_i$ for each $i=1,\dots,n$, we have that $f(x_1,\dots,x_n) \leq f(y_1,\dots,y_n)$.
\end{definition}

\begin{rtheorem}[\thmQueryOneMon]
Given a \cGAP\ $\Pi$, a monotone estimation query ${\cal{Q}}$ over $\Pi$ and a possible answer $a$ of ${\cal{Q}}$, the problem of deciding whether $a = {\tt glb}(poss(\Pi,{\cal{Q}}))$
(resp. $a = {\tt lub}(poss(\Pi,{\cal{Q}}))$) is co-{\bf NP}-hard.~\hfill~$\Box$
\end{rtheorem}
The above theorem says that finding the range answer to a monotone estimation query is computationally hard to compute. Fortunately, the next result suggests that
by considering specific interpretations, we can get lower and upper bounds on the range answer.

\begin{rtheorem}[\thmLowerB]\label{th:LB}
Let $\Pi$ be a VIC$_2$ program and let ${\cal{Q}}= \langle f(\mu_1,\dots,\mu_n), A_1:\mu_1,\dots,$ $A_n:\mu_n \rangle$ be
a monotone estimation query over $\Pi$. Then,
\begin{itemize}
  \item[1.] $f(I(A_1),\dots,I(A_n)) \leq {\tt glb}(poss(\Pi,{\cal{Q}}))$, where $I = {\cal MM}(\Pi^1_{S_{21}}\cup \Pi^2_{S_{12}})$;
  \item[2.] $ {\tt lub}(poss(\Pi,{\cal{Q}}))\leq f(I(A_1),\dots,I(A_n))$, where $I = {\cal MM}(\Pi^1_{S_{12}}\cup \Pi^2_{S_{21}})$.~\hfill~$\Box$
\end{itemize}
\end{rtheorem}

Note that the monotonicity of function $f$ is necessary to prove Theorem~\ref{th:LB} by using Theorem~\ref{maxmin} (also see Example~\ref{ex:18}).
The following result shows that range answers to monotone queries can be computed by looking at the minimal models of certain specific GAPs.

\begin{rtheorem}[\thmGLB]\label{th:GLB}
Let $\Pi$ be a VIC$_2$ program and let ${\cal{Q}} = \langle f(\mu_1,\dots,\mu_n), A_1:\mu_1,\dots,$ $A_n:\mu_n \rangle$ be
a monotone estimation query over $\Pi$. Let $a_1,\dots,a_n$ be the predicate symbols appearing in atoms $A_1,\dots,A_n$.
If $\{a_1,\dots,a_n\} \subseteq Pred_{\Pi}^1$ (resp. $\{a_1,\dots,a_n\} \subseteq Pred_{\Pi}^2$), then
\begin{itemize}
  \item ${\tt glb}(poss(\Pi,{\cal{Q}})) = f(I(A_1),\dots,I(A_n))$, where $I = {\cal MM}(\Pi_{S_{21}})$ (resp. $I = {\cal MM}(\Pi_{S_{12}})$);
  \item ${\tt lub}(poss(\Pi,{\cal{Q}})) = f(I(A_1),\dots,I(A_n))$, where $I = {\cal MM}(\Pi_{S_{12}})$ (resp. $I = {\cal MM}(\Pi_{S_{21}})$).~\hfill~$\Box$
\end{itemize}
\end{rtheorem}

\begin{example}\label{ex:18}
Consider the situation of Example~\ref{ex:17} and assume that $buyAsus$ is choice $1$, while $buyMac$ is choice $2$.
Then, $SE_a = {\cal MM}(\Pi_{S_{12}})$ and $SE_b = {\cal MM}(\Pi_{S_{21}})$. 
By considering the monotone estimation query ${\cal Q} = \langle {\tt sum}(\mu),$\\ $\{buyAsus^D(1):\mu\} \rangle$ we have that 
${\tt glb}(poss(\Pi,{\cal{Q}}))=SE_b(buyAsus^D) = 0.3$ and ${\tt lub}(poss(\Pi,{\cal{Q}}))=SE_a(buyAsus^D) = 0.6$, and this coincide with the range answer to ${\cal Q}$.

If we now consider the non-monotone estimation query ${\cal Q'} = \langle 1-{\tt sum}(\mu),$ \\$\{buyAsus^D(1):\mu\} \rangle$, we have that 
${\tt glb}(poss(\Pi,{\cal{Q}}))\neq 1-SE_b(buyAsus^D) = 0.7$ and ${\tt lub}(poss(\Pi,{\cal{Q}}))$ $\neq 1-SE_a(buyAsus^D) = 0.4$.
\hfill$\Box$
\end{example}

The following result shows that range answers for monotone queries can be computed in polynomial time. In fact,  Theorem~\ref{th:GLB}
tells us that the range answer can be computed by just using the \SEs\ ${\cal MM}(\Pi_{S_{21}})$ and ${\cal MM}(\Pi_{S_{12}})$, and these \SEs\ can be computed in PTIME as shown in Theorem~\ref{th:8.6}. 

\begin{corollary}
Let $\Pi$ be a VIC$_2$ program and let ${\cal{Q}} = \langle f(\mu_1,\dots,\mu_n), A_1:\mu_1,\dots,$ $A_n:\mu_n \rangle$ be
a monotone estimation query over $\Pi$. Let $a_1,\dots,a_n$ be the predicate symbols appearing in atoms $A_1,\dots,A_n$.
If $\{a_1,\dots,a_n\} \subseteq Pred_{\Pi}^1$ (resp. $\{a_1,\dots,a_n\} \subseteq Pred_{\Pi}^2$), then computing the range semantics is in PTIME. 
\end{corollary}

\subsection{Linear Queries}
In this section, we focus on linear queries. These queries can be computed on linear cGAPs which are
not necessarily VIC programs.
Moreover, linear queries can express monotone queries such as ${\tt sum}$ and  ${\tt count}$.

\begin{definition}
An estimation query ${\cal{Q}}= \langle f(\mu_1,\dots,\mu_n), A_1:\mu_1,\dots,A_n:\mu_n \rangle$ over a linear \cGAP\ $\Pi$ is \emph{linear} if
the function $f(\mu_1,\dots,\mu_n)$ is computable in PTIME and
\begin{itemize}
\item $f$ is expressed by a linear function $k + c_1^T \bar{x} + c_2^T \bar{y}$, where
$\bar{x} = [x_{A_1},\dots,x_{A_n}]$ is the vector of variables associated with $\mu_1,\dots,\mu_n$,  $\bar{y} = [y_{1},\dots,y_{m}]$ is a vector of $m \leq pol(n)$ real or integer variables, $k$ is a constant, and $c_1$ and $c_2$ are vectors of constants.
\item $\hat{x}$ and $\hat{y}$ are constrained by a set of linear constraints $C_f$ of size $pol(n)$, i.e. $(\hat{x},\hat{y}) \in C_f$.
\end{itemize}
\end{definition}

\begin{rtheorem}[\thmQueryOneLin]
Given a \cGAP\ $\Pi$, a linear estimation query ${\cal{Q}}$ over $\Pi$ and a possible answer $a$ of ${\cal{Q}}$, the problem of deciding whether $a = {\tt glb}(poss(\Pi,{\cal{Q}}))$
(resp. $a = {\tt lub}(poss(\Pi,{\cal{Q}}))$) is co-{\bf NP}-hard.~\hfill~$\Box$
\end{rtheorem}

The following result shows that linear queries can be computed by solving a mixed integer linear program.
%Observe that, queries where the function $f$ is $Min$, $Max$, $Sum$, or $Count$ are linear queries.

\begin{proposition}\label{prop:rangeAns}
Let $\Pi$ be a linear \cGAP\ and ${\cal{Q}}$ a linear estimation query over $\Pi$, then ${\tt glb}(poss(\Pi,{\cal{Q}}))$ is equal to

\[
\begin{array}{rl}
\min & k + c_1^T \bar{x} + c_2^T \bar{y}\\
\mathbf{subject\ to} &\\
& (\hat{x},\hat{y}) \in C_f\\
& \hat{x} \in ILC(\Pi,\widehat{T})
\end{array}
\]

and ${\tt lub}(poss(\Pi,{\cal{Q}}))$ is equal to

\[
\begin{array}{rl}
\max & k + c_1^T \bar{x} + c_2^T \bar{y}\\
\mathbf{subject\ to} &\\
& (\hat{x},\hat{y}) \in C_f\\
& \hat{x} \in ILC(\Pi,\widehat{T})
\end{array}
\]\hfill$\Box$
\end{proposition}

\begin{example} Consider the cGAP $\Pi$ of Example \ref{exspiega} --- its corresponding set $ILC(\Pi,\widehat{T})$ of constraints is described in Examples~\ref{exspiega1}-\ref{exspiega5}.
Let
${\cal{Q}}_1= \langle \mu_1-\mu_2, \{b_1:\mu_1,b_2:\mu_2 \}\rangle$ be an estimation estimation query.
Then, the greatest lower bound is computed using the following MILP:
\[
\begin{array}{rl}
\min & x_{b_1}-x_{b_2}\\
s.t. &\\
& \{x_{b_1},x_{b_2}\} \in ILC(\Pi,\widehat{T}).
\end{array}
\]
To compute the least upper bound, it is sufficient to change the objective function to $\max \; x_{b_1}-x_{b_2}$.

Moreover,  we can express a count estimation query over the individuals that make a specific choice by considering the \emph{choice} atoms, i.e. the atoms indicating that a player makes a particular choice, and setting the objective function to be the sum  of all the associated variables.

As an other example, consider the estimation query 
${\cal{Q}}_2= \langle {\tt min}(\mu_1,\mu_2),\{ b_1:\mu_1,b_2:\mu_2\} \rangle$.
In this case, the greatest lower bound is computed with the following MILP:
\[
\begin{array}{rl}
\min & y_2\\
\mathbf{subject\ to} &\\
& x_{b_1}+y_1\geq y_2\\
& x_{b_1}\leq y_2\\
& x_{b_2}+(1-y_1)\geq y_2\\
& x_{b_2}\leq y_2\\
& y_1\in {0,1}\\
& y_2\in [0,1]\\
& \{x_{b_1},x_{b_2}\} \in ILC(\Pi,\widehat{T}).
\end{array}
\]
The least upper bound is computed by changing the objective function to $\max \;y_2$. 
The ${\tt max}$ aggregate operator can be expressed similarly as ${\tt min}$.\hfill$\Box$
\end{example}

\section{Experiments}\label{sec:expts}
We ran experiments to check the scalability and accuracy of our framework in predicting real-world election outcomes by considering \emph{Facebook} discussions surrounding the 2013 Italian general election.
All experiments were run on an Intel I7 2.70 GHz machine with 8 GB RAM.

\paragraph*{Dataset} We used a dataset extracted from \emph{Facebook}. The dataset contains information about Italian Facebook users and their Facebook friends, together with all Facebook pages that each user likes. For each Facebook \emph{like} we store the page url,  name and type (e.g. Actor/Director, Public Figure, Community, Political Organization, etc.).
The dataset contains about $65\,000$ users, $84\,000$ friendship relations, and $520\,000$ likes. The exact statistics are reported in Figure~\ref{fig:stat}.

As our dataset was extracted after the elections, it contains a lot of user preferences about the political parties and/or politicians involved in the electoral competition, expressed in terms of likes of pages maintained by political parties and/or politicians. There were three main political forces or alliances involved in the election competition, denoted by $p_1$, $p_2$ and $p_3$.
%expressed in terms of likes to Political Organization or Politician Facebook pages.
%The dataset has been extracted after the Italian general elections that took place in February 2013 to determine the members of the Chamber of Deputies and the Senate of the Republic for the 17th Parliament of the Italian Republic. Then, it contains a lot of user preferences about the Italian political parties and/or politicians involved in the electoral competition,
%expressed in terms of likes to Political Organization or Politician Facebook pages.
%There was four political forces or alliances involved in the election competition: (1) \emph{Italy. Common Good} (denoted by $p_1$) consisting of centre-left parties and with leader Pier Luigi Bersani, (2) \emph{Centre-right coalition} (denoted by $p_2$) with leader Silvio Berlusconi, (3) \emph{Five Star Movement} (denoted by $p_3$) with leader Beppe Grillo, and (4) \emph{With Monti for Italy} (denoted by $p_4$), centrist coalition with leader Mario Monti.

For each user $u$ in our Facebook dataset and for each party $p_i$ who participated in the elections, we assigned a confidence value $\rho(u,p_i) \in [0,1]$, $1 \leq i \leq 3$ and $\sum_{i=1}^3 \rho(u,p_i) = 1$, that expresses how much the user $u$ likes the party $p$, as follows. 

\begin{itemize}
\item First of all, we classified the Facebook pages of type Political Organization or Politician contained in the dataset (1002 pages) into three categories, $p_1$, $p_2$ and $p_3$, according to the content of the page.
%\emph{Italy. Common Good} (resp. \emph{Centre-right coalition}, \emph{Five Star Movement}, \emph{With Monti for Italy}) alliance.
\item Second, for each user $u$ and for each party $p_i$  we counted how many Facebook pages of type $p_i$ she/he likes (we denote this value as $\#likes(u,p_i)$). 
\item Third, the value $\rho(u,p_i)$ is then computed as
$\rho(u,p_i) = \frac{\#likes(u,p_i)}{\sum_{i=1}^3 \#likes(u,p_i)}$.
\item Finally, 
we classified a user $u$ as \emph{supporter} of the party $p_i$, if $p_i$ corresponds to the maximum coefficient $\rho(u,p_i)$.
\end{itemize}

In our experiments we considered  4 competitions:
\begin{enumerate}
  \item $p_2$ vs. $p_3$;
  \item $p_2$ and $p_3$ vs. $p_1$;
  %\item $p_1$ and $p_2$ vs. $p_1$ and $p_3$;
  \item $p_2$ vs. $p_1$;
  \item $p_3$ vs. $p_1$;
\end{enumerate}

For each competition, we constructed 20 (training set, validation set) pairs  of data to use in the experiments. We did this as follows: given the set $U$ of users having at least one \emph{like} to a page of type Political Organization or Politician (a total of $1\,439$ users), and
a value $\delta \in [0,100]$, we randomly select $\delta\%$ of the users in $U$ to be part of the training set, while the remaining $(1-\delta)\%$
of users are part of the validation set. We used $20$, $30$, $40$, $50$, $60$, $70$ and $80$ as values for $\delta$. Of course, our algorithm is then executed over the whole network (65K users).

\begin{figure}[t]
\centering
\begin{tabular}{ll}
\hline\hline
\multicolumn{2}{c}{Facebook dataset}\\ \hline\hline
Vertices   & 64889\\
Edges (Undirected)  & 83752\\
%Number of WCC & 1\\
Vertices in largest WCC & 64889 (1.0)\\
Edges in largest WCC     & 83752 (1.0)\\
%Nodes in largest SCC   &  (0.)\\
%Edges in largest SCC   &  (0.)\\
Average degree & 2.58\\
Density & 3.98\,$10^{-5}$\\
Average clustering coefficient  & 0.465\\
Number of triangles & 33857\\
%Fraction of closed triangles & \\
Diameter (longest shortest path) &  4\\
%90-percentile effective diameter   & \\
\hline
\end{tabular}
%
%\medskip
%
%\begin{tabular}{c}\hline\hline
%Node average over vertex neighbors average utility\\
%\hline\hline
%$p_1$ \,\, 0.101\\
%$p_2$ \,\, 0.257\\
%$p_3$ \,\, 0.076\\
%\hline
%\end{tabular}
\caption{Facebook network statistics.}\label{fig:stat}
\end{figure}

\paragraph*{Diffusion models} We used three different diffusion models in our experiments, The first diffusion model is a kind of cascade model in which the likelihood of a vertex adopting a political preference is the average of the likelihoods of its friends adopting that position.

$$model_1:\
\begin{array}{c}
  choice1(v):\frac{1}{|nbr(v)|}\sum_{u \in nbr(v)} \mu_u \leftarrow \bigwedge_{u \in nbr(v)} choice1(u):\mu_u.\\
  choice2(v):\frac{1}{|nbr(v)|}\sum_{u \in nbr(v)} \mu_u \leftarrow \bigwedge_{u \in nbr(v)} choice2(u):\mu_u.
\end{array}
$$

In the second diffusion model, the likelihood of a vertex adopting a political preference is proportional to the maximal likelihood of one of its friends adopting the same position.

$$model_2:\
\begin{array}{c}
  choice1(v):\max_{u \in nbr(v)} \mu_u \leftarrow \bigwedge_{u \in nbr(v)} choice1(u):\mu_u. \\
  choice2(v):\max_{u \in nbr(v)} \mu_u \leftarrow \bigwedge_{u \in nbr(v)} choice2(u):\mu_u.
\end{array}
$$

The third diffusion model used in our experiments is a  tipping model~\cite{Granovetter73,Schelling78}. 
It checks to see if the sum of the likelihoods of the friends of a vertex adopting a particular political preference exceeds a threshold $\tau$ (the tipping point).  If so, the likelihood of the vertex
adopting the same political preference is the likelihood of the neighbor having the strongest political preference.

$$model_3:\
\begin{array}{r}
  choice1(v):\max_{u \in nbr(v)} \mu_u \leftarrow \bigwedge_{u \in nbr(v)} choice1(u):\mu_u \wedge \\\wedge \sum_{u \in nbr(v)} \mu_u \geq \tau. \\
  choice2(v):\max_{u \in nbr(v)} \mu_u \leftarrow \bigwedge_{u \in nbr(v)} choice2(u):\mu_u \wedge \\ \wedge \sum_{u \in nbr(v)} \mu_u \geq \tau.
\end{array}
$$

\begin{figure}[t]
     \centering
     \includegraphics[width=7cm]{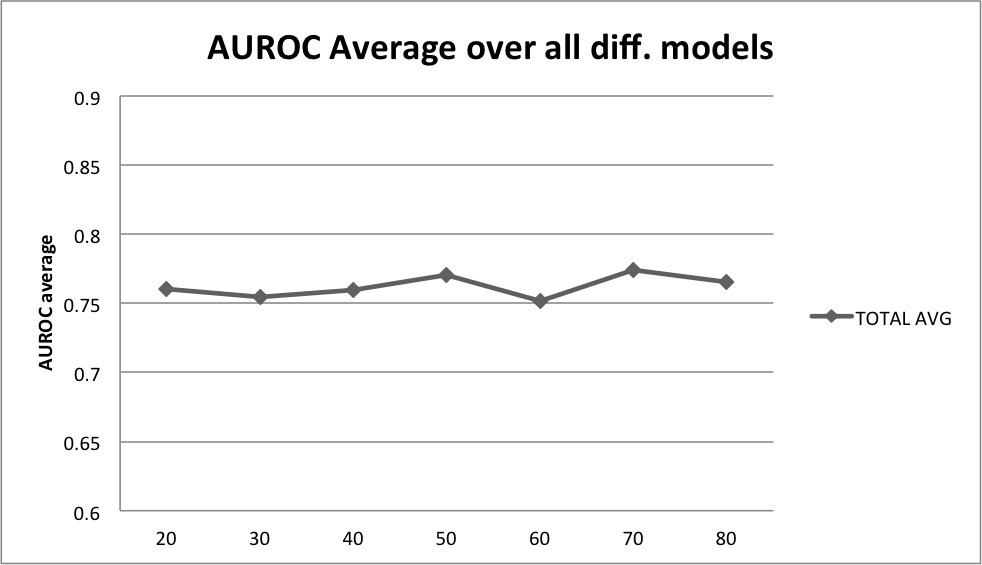}
     \caption{Average AUROC ($y$-axis) as we vary the size of the training set ($x$-axis) from 20\% to 80\% of the entire Facebook data set.}\label{fig:auroc-1}
      \end{figure}

For each competition-diffusion model pair, we  computed the maximal model ($M_1$) and the minimal model ($M_2$) using 
Algorithm \ref{SEalgo}.
We assigned  real utility values to users in the training set, while users in the validation set were assigned a 0 utility value
 (because we will use our model to predict which of the two political orientations these users prefer). The real utility values are computed by taking into account the values $\rho(u,p_i)$. For instance, for the second competition the utility of vertex $u$ for $p_2$ and $p_3$ is
$u_1(u) = \frac{\rho(u,p_2)+\rho(u,p_3)}{\rho(u,p_1)+\rho(u,p_2)+\rho(u,p_3)}$, while the utility of $u$ for $p_1$ is $u_2(u) = \frac{\rho(u,p_1)}{\rho(u,p_1)+\rho(u,p_2)+\rho(u,p_3)}$.

%\paragraph*{Accuracy} Let $T$ be the set of vertices $v$ in the test set, then we denote by $c(v)$ the choice of the vertex $v$ predicted by our framework.
%We have that
%$$c(v) =
%\begin{cases}
%1 & if \ v \ does \ the \ choice \ 1 \ both \ in \ M_1 \ and \ M_2,\\
%2 & if \ v \ does \ the \ choice \ 2 \ both \ in \ M_1 \ and \ M_2,\\
%uncert & if \ v \ does \ the \ choice \ 1  \ in \ M_1 \ and\ the\ choice\ 2 \ M_2\ or\ viceversa.\\
%\end{cases}
%$$
%
%Then, in order to take into account the fact that some vertices are predicted as uncertain ($c(v)=uncert$) as the utilities of the possible choices are very close,
%we defined the function $a(v)$ as follows
%$$a(v) =
%\begin{cases}
%1 & if \ c(v)=1\ \wedge u_1 \geq u_2,\\
%1 & if \ c(v)=2\ \wedge u_2 \geq u_1,\\
%0 & if \ c(v)=1\ \wedge u_1 < u_2,\\
%0 & if \ c(v)=2\ \wedge u_2 < u_1,\\
%1-|u_1-u_2| & if\ c(v)=uncert.
%\end{cases}
%$$
%where that $|u_1-u_2|$ expresses how much the two utility values $u_1$ and $u_2$ are different.
%%
%Finally, the accuracy is computed as
%
%$$accuracy = \frac{\sum_{v \in T} a(v)}{|T|}$$

\begin{figure}[t]
     \centering
     \includegraphics[width=7cm]{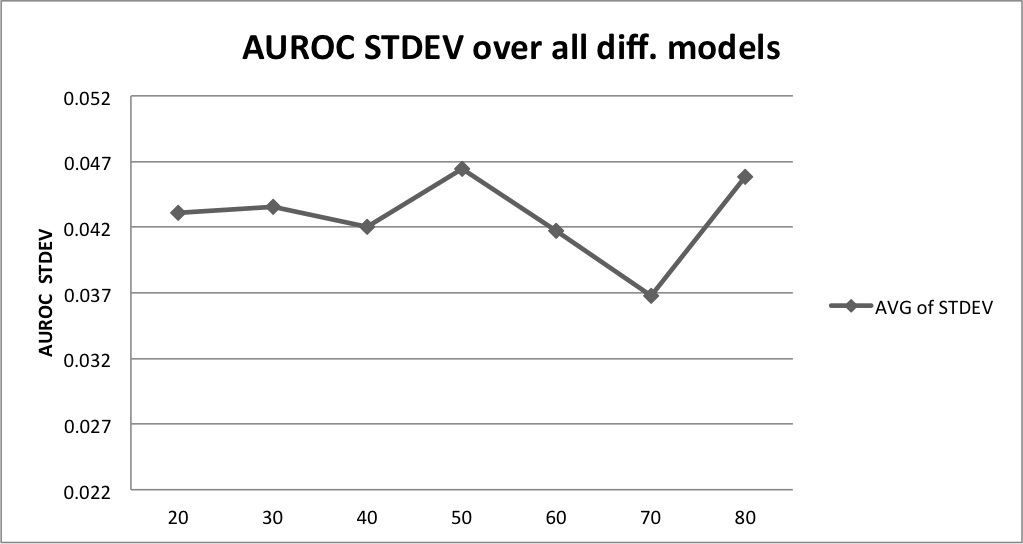}
     \caption{Standard deviation of AUROC ($y$-axis) as we vary the size of the training set ($x$-axis) from 20\% to 80\% of the entire Facebook data set.}\label{fig:auroc-2}
      \end{figure}

\paragraph*{Area under the ROC curve} In order to evaluate our model, we built a threshold classifier by using bounds over the vertices' utility values returned by the models $M_1$ and $M_2$. For each vertex $i$, we have a lower bound $L_1(i)$ (resp. $L_2(i)$) and an upper bound $U_1(i)$ (resp. $U_2(i)$) over the utility of choice $1$ (resp. choice $2$) for vertex $i$.
%These values are obtained as follows:
%\begin{itemize}
%\item $L_1(i) = $
%\item $U_1(i) = $
%\item $L_2(i) = $
%\item $U_2(i) = $
%\end{itemize}
For each vertex $i$ we computed the value $f(i) = \frac{L_1(i)+U_1(i)}{2} - \frac{L_2(i)+U_2(i)}{2}$. The first term denotes the utility of preference 1 (choice 1) for vertex $i$, computed as the average of the lower and upper bound estimates of that utility. The second term represents the same concept for the second choice.  $f(i)$ captures the difference between these utilities. When this difference exceeds a given threshold $\tau$, the vertex $i$ goes with the first choice of political party. If not, it goes with the second choice.

We computed the ROC curve by varying the threshold $\tau$, and used the area under the ROC curve to measure the accuracy of our model.

Figure~\ref{fig:auroc-1} shows a graph of the average AUROC as we vary the size of the training set from 20\% of the overall data set to 80\% in steps of 10\%.
For each value of this size, we randomly selected a data set of that size from our Facebook data in 20 ways. Then, as mentioned above, another 4500 combinations of parameters were considered, making a total of 90,000 experimental settings for each data set size. We have a total of 8 data set sizes, making 720K runs in all in our experiments.

For each data set size, Figure~\ref{fig:auroc-1} shows the average AUROC we derived. We see that on average, the AUROC varies between 0.75 and 0.78 which is quite a narrow band.
%(fluctuations may appear highly magnified in Figure~\ref{fig:auroc-1} because the $y$-axis is focused between 0.7 and 0.74). 
Recall that an AUROC of 0.5 denotes random guessing and hence these AUROCs show strong predictive power.  Moreover, the predictive power seems relatively flat as we vary the size of the training set from 20\% to 80\%, varying by just about 3 percentage points overall, which indicates that we can get good predictive accuracy even without large training sets.

Appendix~\ref{app:tables} shows 9 tables that provide more detail, on a competition by competition level, of the experimental accuracy summarized in Figure~\ref{fig:auroc-1}. Each table corresponds to one way of assigning a diffusion model to how each of the two political preferences spreads, leading to $(3\times 3)=9$ tables in all.
Each of these tables shows the diffusion models used for the spread of the two political preferences, as well as the AUROCs obtained with no perturbation at all.
%, and with vertex perturbations and edge perturbations. 
Each table shows these values for each of the 4 competitions as we vary the training set from 20 to 80\% of the total Facebook data set.

Figure~\ref{fig:auroc-2} shows the standard deviation of the AUROCs we obtained.  This figure shows an important trend, namely that the standard deviation stays small, under 0.05, even as the size of the training set varies from 20\% to 80\%.

Table~\ref{tab:maxAuroc} reports, for each competition and for each training set, the max AUROC obtained, the corresponding diffusion model, and the running time. 
%EXPLAIN WHY THE FIRST ONE IS BAD
Observe that the AUROCs obtained are all good except in the case of the first competition. This one case is anomalous because amongst the users in the set $U$ (i.e. having at least one like to a political party), the percentage of users supporting party $p_2$ (resp. $p_3$) is 44\% (resp. 37\%).  However, each user $u \in U$ supporting $p_3$ has, on average, 59\% of friends supporting $p_2$ and 16\% supporting $p_3$.  On the other hand, each user $u \in U$ supporting $p_2$ has, on average, 70\% of friends supporting $p_2$ and just 6\% supporting $p_3$. It follows that the users in $U$ that actually support $p_3$ will be denoted as users supporting $p_2$ by the diffusion models.  
 
\paragraph*{Network perturbation} In order to ensure that our experimental results are
 robust to (edge) noise in the network, we perturbed all the best scenarios of our Facebook network shown in Table~\ref{tab:maxAuroc} in two ways.

\begin{figure}
        \centering
        \begin{subfigure}[b]{0.45\textwidth}
                \includegraphics[width=\textwidth]{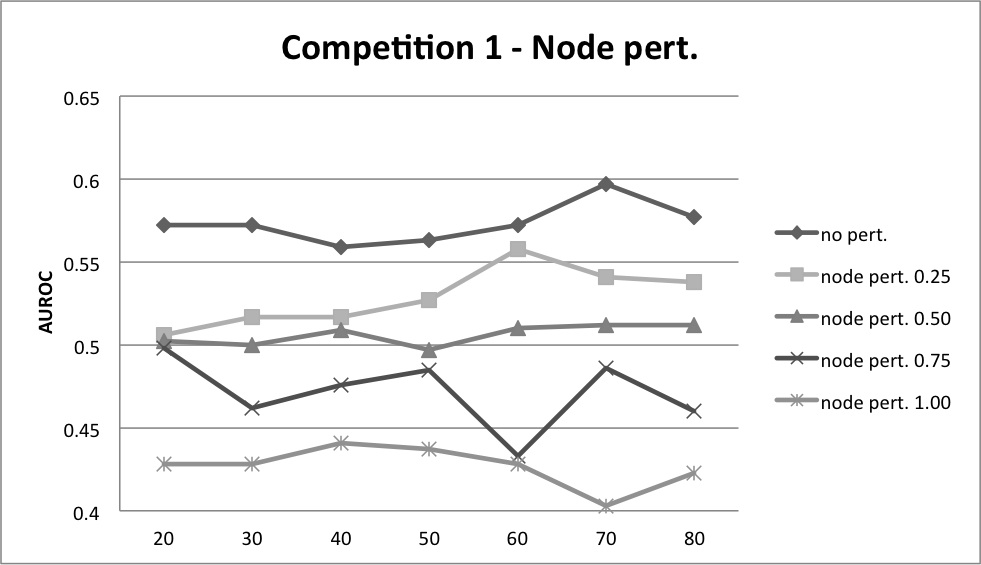}
                \caption{}
                \label{fig:N1}
        \end{subfigure}%
        ~ %add desired spacing between images, e. g. ~, \quad, \qquad, \hfill etc.
        \begin{subfigure}[b]{0.45\textwidth}
                \includegraphics[width=\textwidth]{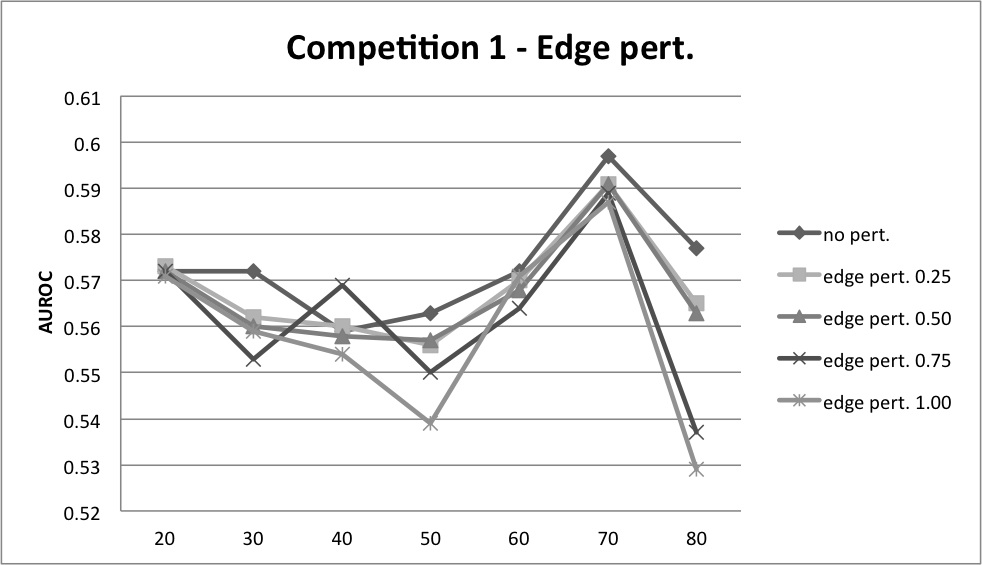}
                \caption{}
                \label{fig:tiger}
        \end{subfigure}
        \quad
        ~ %add desired spacing between images, e. g. ~, \quad, \qquad, \hfill etc.
        \begin{subfigure}[b]{0.45\textwidth}
                \includegraphics[width=\textwidth]{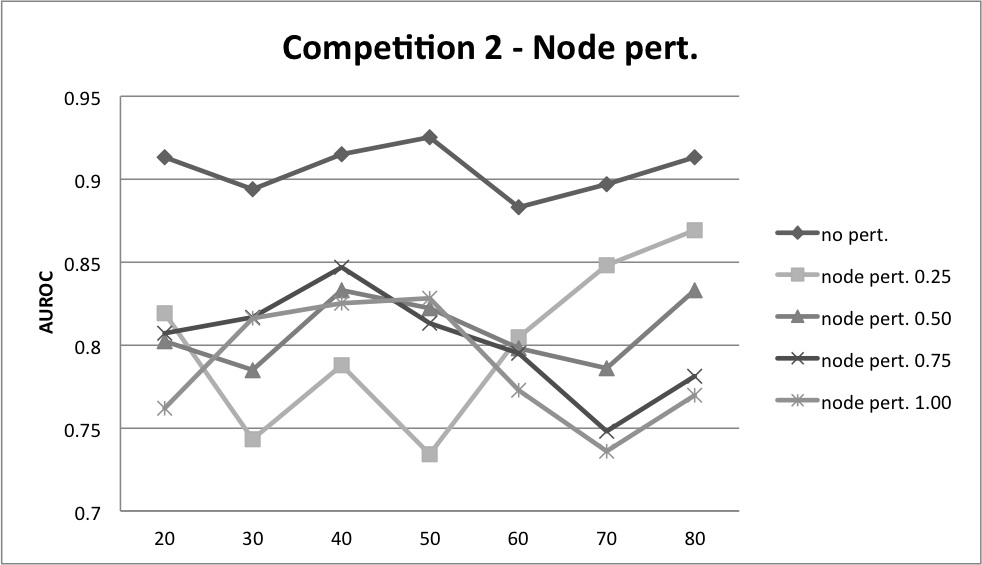}
                \caption{}
                \label{fig:mouse}
        \end{subfigure}
        \begin{subfigure}[b]{0.45\textwidth}
                \includegraphics[width=\textwidth]{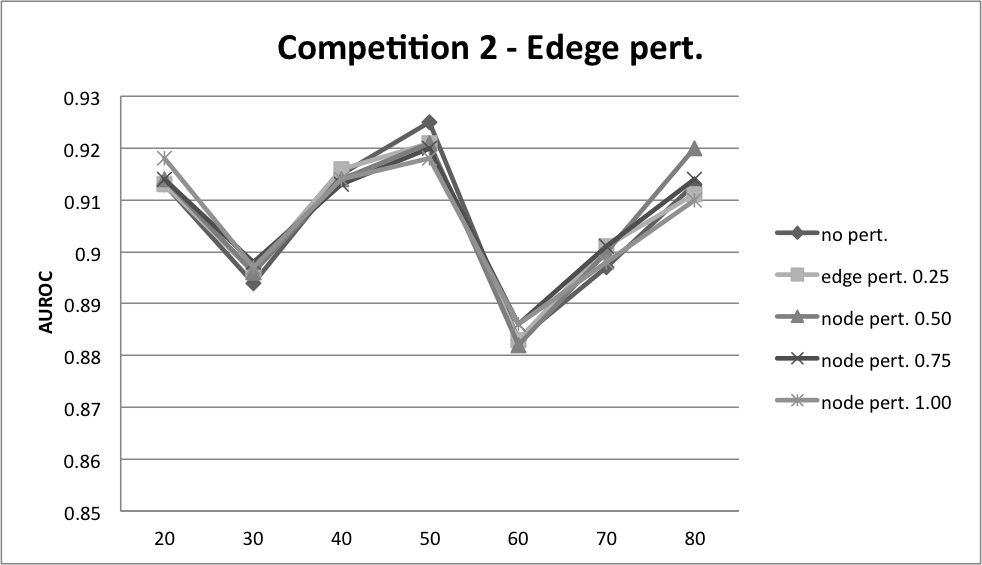}
                \caption{}
                \label{fig:mouse}
        \end{subfigure}
        \quad
        ~ %add desired spacing between images, e. g. ~, \quad, \qquad, \hfill etc.
        \begin{subfigure}[b]{0.45\textwidth}
                \includegraphics[width=\textwidth]{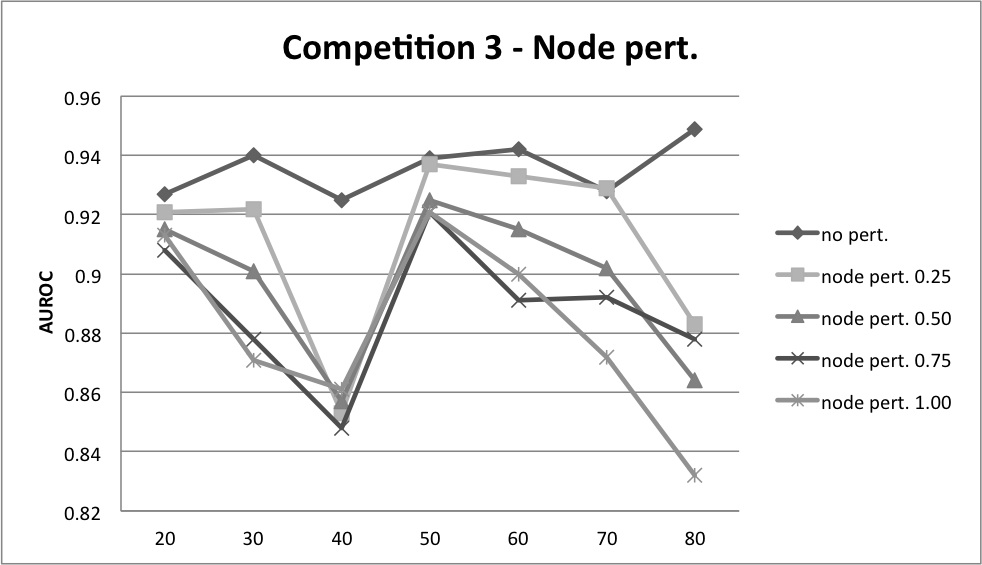}
                \caption{}
                \label{fig:mouse}
        \end{subfigure}
        \begin{subfigure}[b]{0.45\textwidth}
                \includegraphics[width=\textwidth]{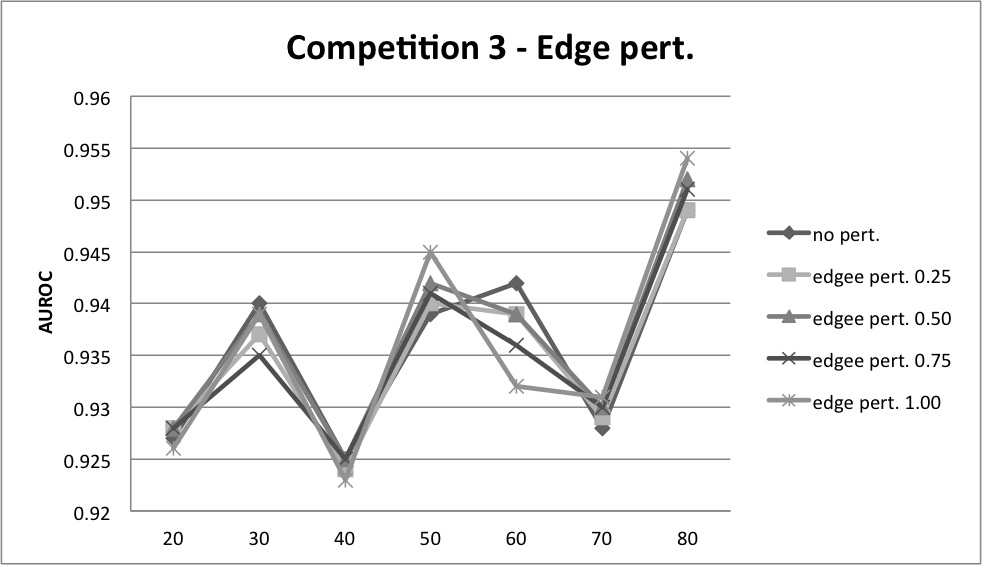}
                \caption{}
                \label{fig:mouse}
        \end{subfigure}
        \quad
        ~ %add desired spacing between images, e. g. ~, \quad, \qquad, \hfill etc.
        \begin{subfigure}[b]{0.45\textwidth}
                \includegraphics[width=\textwidth]{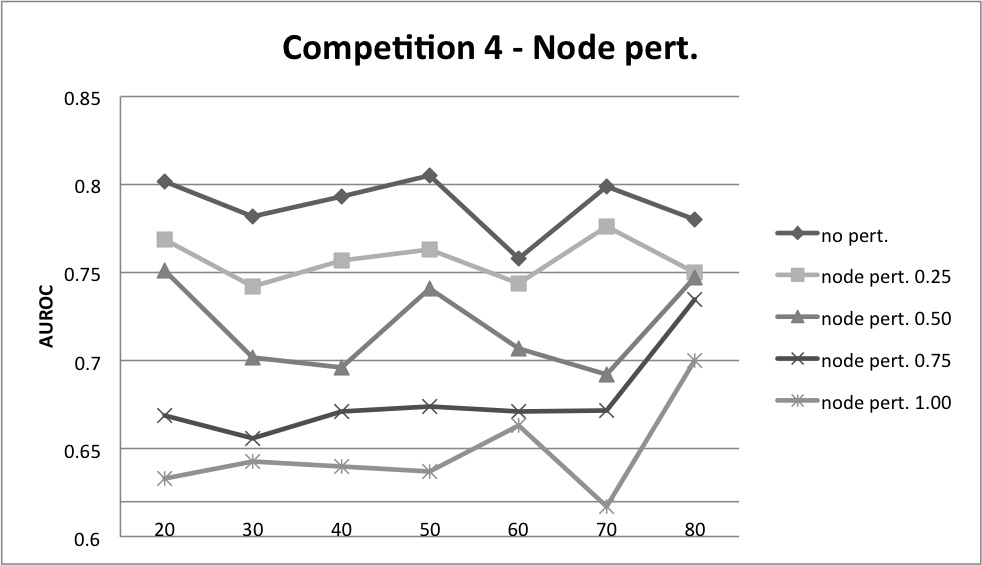}
                \caption{}
                \label{fig:mouse}
        \end{subfigure}
        \begin{subfigure}[b]{0.45\textwidth}
                \includegraphics[width=\textwidth]{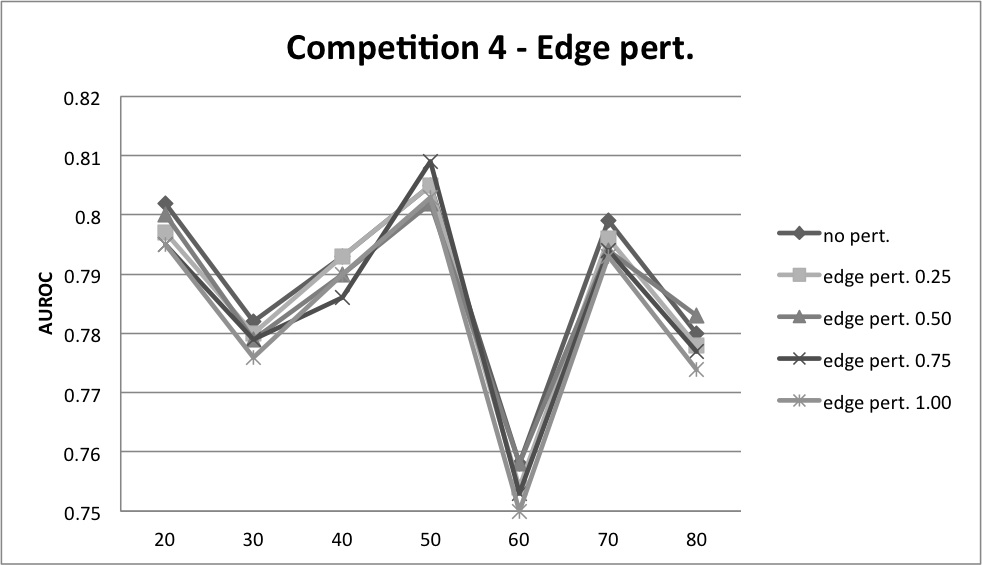}
                \caption{}
                \label{fig:mouse}
        \end{subfigure}
        \caption{Network perturbation.}\label{fig:perturbation}
\end{figure}

\emph{Node Perturbation.}
We perturbed vertices by exchanging the utilities of the two possible choices that a vertex might make in the training set with probability $p$. 
%For each configuration of competition (5 choices) and each pair of diffusion models adopted in each competition ($3\times 3=9$ choices), we tested with 100 perturbed samples of the network, leading to a total of 4500 choices.
For each case shown in Table~\ref{tab:maxAuroc}, we did this by varying  $p\in \{$ 25\%, 50\%, 75\%, 100\% $\}$.
The results are shown in Figure~\ref{fig:perturbation} (first column of plots). In general, for all the competitions we observe the same trend:  increasing $p$ yields a lower AUROC. This is because our framework infers the political orientation of all the vertices in the network starting from  vertices which like at least one political party in the training set.
If this initial knowledge is wrong, then the AUROC is lower than the case without vertex perturbation. If the premise is wrong, then so is the conclusion.

\emph{Edge Perturbation.}
The second method guesses a set $GE$ of edges where $|GE|$ is a percentage $p$ of the total number of edges in the graph. Think of $GE$ as edges that are wrongly represented in the graph. If an edge in $GE$ is in the graph, it must be removed. If an edge in $GE$ is not in the graph, it is inserted into the graph.
Here too, we perturbed all the cases shown in Table~\ref{tab:maxAuroc} with the following values of probability $p$: 25\%, 50\%, 75\%, and 100\%. An edge perturbation with $p=$100\% means that the number of the edges is at most doubled.  
The results are shown on Figure~\ref{fig:perturbation} (second column of plots). In this case, our framework is robust to the noise introduced, since we can observe very low variations between the AUROCs in the perturbed cases and the ones without perturbation.

\begin{table}[t]
\centering
{\tiny
\begin{tabular}{|c|c|c|c|c|c|}\hline
mod1 & mod2 & $\%$ Training set & Comp & AUROC& Time (ms)\\\hline
2 & 3 & 20 & 1 & 0.572& 26908\\
3 & 2 & 30 & 1 & 0.572 & 12276\\
3 & 3 & 40 & 1 & 0.559 & 27202\\
3 & 3 & 50 & 1 & 0.563& 27089\\
1 & 3 & 60 & 1 & 0.572 & 9454\\
3 & 2 & 70 & 1 & 0.597 & 11720\\
3 & 3 & 80 & 1 & 0.577& 27040\\\hline
1 & 2 & 20 & 2 & 0.913 & 60823\\
1 & 2 & 30 & 2 & 0.894& 58445\\
1 & 2 & 40 & 2 & 0.915 & 53038\\
1 & 2 & 50 & 2 & 0.925 & 54272\\
1 & 2 & 60 & 2 & 0.883 & 49449\\
1 & 1 & 70 & 2 & 0.897 & 44034\\
1 & 3 & 80 & 2 & 0.913 & 9331\\\hline
1 & 3 & 20 & 4 & 0.927 & 9258\\
1 & 3& 30 & 4 & 0.94 & 9421\\
1 & 3 & 40 & 4 & 0.925 & 9518\\
1 & 1 & 50 & 4 & 0.939 & 36696\\
1 & 1 & 60 & 4 & 0.942 & 35688\\
1 & 3 & 70 & 4 & 0.928 & 9401\\
2 & 2 & 80 & 4 & 0.949 & 35790\\\hline
1 & 3 & 20 & 5 & 0.802 & 9219\\
1 & 3 & 30 & 5 & 0.782 & 9300\\
1 & 3 & 40 & 5 & 0.793 & 9681\\
1 & 3 & 50 & 5 & 0.805& 9632\\
1 & 3 & 60 & 5 & 0.758 & 9577\\
1 & 3 & 70 & 5 & 0.799 & 9399\\
1 & 3 & 80 & 5 & 0.78 & 9527\\
\hline
\end{tabular}
\caption{Maximum AUROC per each competition and per each size of training set.}\label{tab:maxAuroc}
}
\end{table}

\section{Related Work}
To the best of
our knowledge, this paper presents the first game-theoretic framework for competitive diffusion which scales to large social networks and which has been proven to have high accuracy.  

\cite{kleinberg} proposes a framework for diffusion which is less general than the one presented here.  The authors study a different problem than that presented in this paper - namely the problem of identifying the``most influential vertices'' in a network.  \cite{kleinberg} does not address either competitive diffusion or scaling to large social networks.  
Moreover, all diffusion models considered by \cite{kleinberg} must all conform to a submodularity property.  A key difference between the approach of \cite{kleinberg} and the frameworks here and in \cite{SNDOP_query} is that \cite{kleinberg} does not consider different edge and vertex properties.  Additionally, there are many diffusion models in which submodularity does not hold --- a simple diffusion model is presented in \cite{snops-iclp} where submodularity does not hold - based on a property spreading via multiple diffusion rules.  This argument easily applies for competitive diffusion models as well.  It is important to note that \cite{kleinberg} is a generalization of many important models in economics and social science such as \cite{Gran78,jy05,libai} - which can be represented with our logic-programming based framework as we showed in \cite{SNDOP_query}.

\cite{kleinberg} is extended to a competive scenario in \cite{carnes07}. However, \cite{carnes07} only allows one competitor to actively diffuse while all others must be static. In our work, all competitors are active at the same time.
They extend the independent cascade model, and propose two models describing how two technologies simultaneously diffuse over the network. In the first model (distance based), the vertex $u$ influences a vertex $v$ if they are connected and their distance in the network is small. In the second model (wave propagation) a vertex copies the adoption of a neighboring vertex randomly chosen from the set of its  neighbors that are closest to the seeds. They show that for the two proposed diffusion models the decision versions are NP-hard, and the corresponding influence functions are non-negative, monotone, and sub-modular. A
small dataset (8K vertices, 461 connected components) was used for experiments.
From a game-theoretic perspective, the paper found a best response to the first player's move in a Stackelberg game. In contrast, our work is applicable to any diffusion models, not just three, we prove a host of complexity results and algorithms that take various aggregate functions into account, and we do not require submodularity. We also test our algorithms on a much bigger data set.

\cite{kostka08} provides a theoretical treatment of a problem similar to \cite{carnes07} with respect to the spread of rumors. 
They addressed the problem as a 2-player-game where the players are the competitors or rumors (in contrast, in our case the players are the vertices). The first player starts by choosing its set of seeds, then the second player makes his choice of seeds (the two seeds sets are different). Rumors then propagate on the network according to specific propagation models. The payoff of a player is computed after the diffusion has terminated and equals the number of vertices that believes in the rumor corresponding to the player. They show that computing the optimal strategy for both players is NP-complete, as well as determining an approximated solution for the first player. They analyzed several heuristics and showed that being the first to decide is not always advantageous. Their framework did not include an implementation. 

The well-known work of \cite{lieberman_evolutionary_2005} in biology presents a competitive diffusion model that considers the spread of a mutant gene in a population.  As the mutant gene is attempting to replace (and can be replaced) by ``residents'' - the model is essentially competitive. The authors view the propagation of the mutants and residents as a stochastic process - which is similar to the long-studied ``Moran process''~\cite{moran58} in evolutionary biology - a non-graphical model which their work generalizes.  It is important to note that the model of  \cite{lieberman_evolutionary_2005} so far has not been used in real-world algorithms and applications, but rather has been used as a theoretical basis for drawing conclusions on \textit{fixation probability} - the probability that a single mutant overtakes a population as time progresses indefinitely.  Unlike this work, their model only has two competitors (mutants and residents), does not consider multiple edge labels, or vertex properties beyond mutant/resident.  We also note that \cite{lieberman_evolutionary_2005} has influenced work in statistical physics and interacting particle systems such as \cite{sood08} that uses basically the same intuition - and differs from this work in the same manner as \cite{lieberman_evolutionary_2005} does. In contrast to these efforts,  
we allow more than two competitors, allow vertices and edges to have property labels, and in addition, our framework of \choicegaps\ allows us to express a huge variety of diffusion models whereas these other papers focus on just one type of diffusion model. We have also proved a suite of complexity results
and implemented scalable algorithms that avoid costly Monte-Carlo style simulations typically used for such models (see \cite{DBLP:conf/ijcai/RaedtKT07}).

\cite{BroechelerSS10} presented a logic-based competitive diffusion model in which they induced a probability distribution over the space of models of a annotated logic program and hypothesized that the most likely ``model'' was the one that was likely to happen in practice. This one model of the annotated logic program was used to make forecasts (e.g. if the model said more people would adopt choice 1 instead of choice 2, then that is what would happen). However, no accuracy results were presented. One flaw with this is that it is possible that the most probable model has probability 5\% and suggests that more people would adopt choice 1, while the remaining  models (carrying 95\% probability of occurring) suggest they would adopt choice 2.
Moreover, even on networks of under 10K vertices, the algorithm tooks many, many hours to compute. In contrast, the results of this paper show far greater scalability as well as strong accuracy results.

This paper builds upon our previous work \cite{SNDOP_query} in which we laid out an annotated-logic programming framework for diffusion in networks.  However, this paper differs in several key aspects.  
The approach in \cite{SNDOP_query} does not consider competitive diffusion models. As a consequence, it utilizes a less expressive semantics than the combination of choice logic programs and annotated logic programs used in this paper. It contains no results on equilibria, and does not provide  any accuracy results.
Second, \cite{SNDOP_query} addresses a different problem --- that of determining which vertices in the network will cause a property to spread to a maximal extend (w.r.t. a complex aggregate).
The problem considered here is determine how competing properties will diffuse.

The study of epidemiology is another important field where network diffusion has been intensively studied.  In \cite{snops-iclp}, we showed that the well-known SIR (\textit{susceptible, infectious, removed}) model of disease spread~\cite{anderson79} can be represented in our framework - the same intuitions can be used for other disease modes such as SIS, for example.  
Moreover, \cite{SNDOP_query} shows that other well known diffusion models like the Jackson-Yariv tipping model for product adoption~\cite{jy05}, the cascade model that captures favoriting of photos
on Flickr~\cite{ChaMisloveGummadi_2009}, as well as homophilic models~\cite{aral09} can all be expressed within the notion of a GAP. As \choicegaps\ extend GAPs, they can also be expressed within the \choicegap\ paradigm presented in this paper.

Further works about annotation and probabilistic   logic can be found in  \cite{DBLP:journals/jar/KiferL92,DBLP:conf/slp/SwiftW94a,DBLP:journals/ai/Kern-IsbernerL04,DBLP:journals/ijar/LukasiewiczS09,DBLP:journals/jiis/LeachL96,DBLP:series/sfsc/LukasiewiczS13,DBLP:journals/logcom/Lu96}, while for other works about non-determinism and choice construct in deductive systems see \cite{SaccaZ91,DBLP:journals/logcom/SaccaZ97,GrecoSZ95,GrecoZ98,DBLP:journals/jcss/GiannottiPZ01,DBLP:conf/dood/GiannottiPSZ91,DBLP:journals/amai/PedreschiR04}.

\section{Conclusion}
In  real world social networks, multiple diffusive phenomena are competing for the attention of the same individual.  For instance, multiple competing Presidential candidates are competing for the attention of social network users. Likewise, multiple phone companies (e.g. AT\& T, Verizon, Sprint) are competing for the attention of social network users. Most past work (with a couple of exceptions noted earlier~\cite{kostka08,BroechelerSS10,gandalf13}) have focused on diffusion in a simple setting where only one phenomenon is diffusing at a time

In this paper, we take the problem of competing diffusions ``head on''. We provide a very general framework called \choicegaps\ in which a mix of generalized annotated programs (GAPs) due to Kifer and Subrahmanian~\cite{ks92} are neatly combined with Choice Logic Programs due to Sacc{\`a} and Zaniolo~\cite{SacZan90}.  \choicegaps\ extend the work of \cite{SNDOP_query} where no competition was allowed and a single diffusion was spreading. As GAPs were shown in \cite{SNDOP_query} to be capable of expressing a range of diffusion models including a variety of tipping models, a variety of cascade models, the SIR model of disease spread, the Jackson-Yariv model for product diffusion, a popular Flickr diffusion model~\cite{ChaMisloveGummadi_2009} as well as homophilic models~\cite{aral09}, the \choicegap\ framework can express all such diffusion models. 

Using \choicegaps, we build a game-theoretic framework in which every  social network user is considered to be a player --- thus the resulting game consists of a very large number of players.  We show that certain models of the resulting logic programs can be thought of as ``strong equilibria'' models and that these models have very nice properties, similar to Nash equilibria (whose direct use would require making some unrealistic assumptions). We prove a host of results on the problem of identifying the existence of such equilibria under a variety of settings, and provide algorithms to compute them (or determine they don't exist) that are provably sound and complete. Because finding such strong equilibria can be NP-hard, we identify a tractable class of \cGAP s called $VIC_2$ programs that have two nice properties. First, strong equilibria are guaranteed to exist, and second, they are guaranteed to be computable in polynomial time. 
%Moreoever, most existing diffusion models are $VIC_2$.
We also show that monotone and linear estimation queries can be polynomially computed. Hence, 
as the annotation functions in the heads of \cGAP\ rules are monotonic in the case of the Flickr diffusion model
\cite{mislove09}, the Jackson-Yariv model\cite{jy05}, and homophilic models \cite{aral09}, it is easy to see that these models all can be polynomially computed.

This important result allows us to implement the \choicegap\ framework for realistic applications.

We round out the paper with a prototype implementation of the \choicegap\ framework in which we consider a real-world competitive diffusion situation --- the 2013 Italian election which involved 3 major competing parties (and candidates for Prime Minister). Using a data set that we extracted from Facebook, we show that our framework and algorithm works well in practice, seamlessly handling social networks with about 65K users and 85K edges. Our ability to predict the number of those who like various political parties, even with generic diffusion models from the literature, has an AUROC of 0.76 on average across all experiments done in this paper. Learning diffusion models is not the topic of this paper --- but we suspect that with learned diffusion models, this AUROC would only go up. 

Nonetheless, this work merely represents the tip of the iceberg. There is much additional work that can be done. For instance, it is possible that competitive diffusion is not completely exclusive. A person might say that he likes candidate A's policy on immigration but likes candidate B's policies on jobs. In the case of phone choices, a person might actually own both an AT\&T and a Verizon phone. This suggests that a more general version of competitive diffusion is a constrained choice diffusion where a person is constrained in the number of choices he makes. 
Another major challenge is that of scalability. Though reasoning about competitive diffusion with a 65K vertex network is an order of magnitude bigger than what we have seen in past work~\cite{kleinberg}, it is not yet scalable to real social networks which may have millions of users. A possible strategy to explore this scalability is to coarsen networks along the lines reported in 
\cite{DBLP:conf/kdd/PurohitPKZS14} which shows methods to scale influence maximization to networks with over 30M edges.

% Acknowledgments
\begin{acks}
We are grateful to ARO
for funding this work under grants W911NF11103 and
W911NF09102.
\end{acks}

\bibliographystyle{acmtrans}
\bibliography{competitive,network-comp-diff,network}

% Electronic Appendix
\newpage
\appendix

\section{Detailed Experimental Breakdown}\label{app:tables}

\begin{center}
{\tiny
\begin{tabular}{|c|c|c|c|c|c|}\hline
mod1 & mod2 & $\%$ Training set & Comp & AUROC & Time (ms)\\
 &  &  &  &  &  \\\hline
1 & 1 & 20 & 1 & 0.553 & 48239 \\
1 & 1 & 30 & 1 & 0.549 & 41052 \\
1 & 1 & 40 & 1 & 0.525 & 39886 \\
1 & 1 & 50 & 1 & 0.525 & 42201 \\
1 & 1 & 60 & 1 & 0.566 & 41203 \\
1 & 1 & 70 & 1 & 0.574 & 39942 \\
1 & 1 & 80 & 1 & 0.52 & 38657 \\\hline
1 & 1 & 20 & 2 & 0.895 & 44957 \\
1 & 1 & 30 & 2 & 0.891 & 44725 \\
1 & 1 & 40 & 2 & 0.897 & 43213 \\
1 & 1 & 50 & 2 & 0.917 & 44665 \\
1 & 1 & 60 & 2 & 0.88 & 41384 \\
1 & 1 & 70 & 2 & 0.897 & 44034 \\
1 & 1 & 80 & 2 & 0.912 & 40321\\\hline
1 & 1 & 20 & 3 & 0.926 & 35758 \\
1 & 1 & 30 & 3 & 0.938 & 38353 \\
1 & 1 & 40 & 3 & 0.924 & 35286 \\
1 & 1 & 50 & 3 & 0.939 & 36696\\
1 & 1 & 60 & 3 & 0.942 & 35688\\
1 & 1 & 70 & 3 & 0.927 & 34824\\
1 & 1 & 80 & 3 & 0.943 & 35621\\\hline
1 & 1 & 20 & 4 & 0.799 & 39085\\
1 & 1 & 30 & 4 & 0.776 & 38115 \\
1 & 1 & 40 & 4 & 0.784 & 36340 \\
1 & 1 & 50 & 4 & 0.797 & 36592 \\
1 & 1 & 60 & 4 & 0.754 & 35984 \\
1 & 1 & 70 & 4 & 0.791 & 37878 \\
1 & 1 & 80 & 4 & 0.777 & 36315 \\\hline
\end{tabular}
}
\end{center}

\begin{center}
{\tiny
\begin{tabular}{|c|c|c|c|c|c|}\hline
mod1 & mod2 & $\%$ Training set & Comp & AUROC & Time (ms)\\
 &  &  &  &  &  \\\hline
2 & 2 & 20 & 1 & 0.546 & 46052  \\
2 & 2 & 30 & 1 & 0.557 & 39867 \\
2 & 2 & 40 & 1 & 0.541 & 39858 \\
2 & 2 & 50 & 1 & 0.528 & 40248 \\
2 & 2 & 60 & 1 & 0.555 & 38539 \\
2 & 2 & 70 & 1 & 0.595 & 38242 \\
2 & 2 & 80 & 1 & 0.518 & 36198 \\\hline
2 & 2 & 20 & 2 & 0.906 & 42596 \\
2 & 2 & 30 & 2 & 0.872 & 41465 \\
2 & 2 & 40 & 2 & 0.859 & 40342 \\
2 & 2 & 50 & 2 & 0.897 & 41834 \\
2 & 2 & 60 & 2 & 0.875 & 38450 \\
2 & 2 & 70 & 2 & 0.879 & 41731 \\
2 & 2 & 80 & 2 & 0.901 & 37916 \\\hline
2 & 2 & 20 & 3 & 0.892 & 40180 \\
2 & 2 & 30 & 3 & 0.887 & 41103 \\
2 & 2 & 40 & 3 & 0.912 & 36643 \\
2 & 2 & 50 & 3 & 0.937 & 39429 \\
2 & 2 & 60 & 3 & 0.917 & 37722 \\
2 & 2 & 70 & 3 & 0.916 & 37944 \\
2 & 2 & 80 & 3 & 0.949 & 35790 \\\hline
2 & 2 & 20 & 4 & 0.711 & 37118 \\
2 & 2 & 30 & 4 & 0.724 & 38111 \\
2 & 2 & 40 & 4 & 0.752 & 35863 \\
2 & 2 & 50 & 4 & 0.784 & 37085 \\
2 & 2 & 60 & 4 & 0.702 & 37529 \\
2 & 2 & 70 & 4 & 0.753 & 39792 \\
2 & 2 & 80 & 4 & 0.752 & 37351 \\\hline
\end{tabular}
}
\end{center}

\begin{center}
{\tiny
\begin{tabular}{|c|c|c|c|c|c|}\hline
mod1 & mod2 & $\%$ Training set & Comp & AUROC & Time (ms)\\
 &  &  &  &  &  \\\hline
3 & 3 & 20 & 1 & 0.526 & 26896  \\
3 & 3 & 30 & 1 & 0.558 & 27016 \\
3 & 3 & 40 & 1 & 0.559 & 27202 \\
3 & 3 & 50 & 1 & 0.563 & 27089 \\
3 & 3 & 60 & 1 & 0.531 & 27041 \\
3 & 3 & 70 & 1 & 0.554 & 26983 \\
3 & 3 & 80 & 1 & 0.577 & 27040 \\\hline
3 & 3 & 20 & 2 & 0.678 & 26956 \\
3 & 3 & 30 & 2 & 0.664 & 27059 \\
3 & 3 & 40 & 2 & 0.64 & 27173 \\
3 & 3 & 50 & 2 & 0.639 & 27104 \\
3 & 3 & 60 & 2 & 0.612 & 27421 \\
3 & 3 & 70 & 2 & 0.688 & 26999 \\
3 & 3 & 80 & 2 & 0.631 & 26953 \\\hline
3 & 3 & 20 & 3 & 0.735 & 26951 \\
3 & 3 & 30 & 3 & 0.68 & 27054 \\
3 & 3 & 40 & 3 & 0.701 & 27112 \\
3 & 3 & 50 & 3 & 0.723 & 27219 \\
3 & 3 & 60 & 3 & 0.737 & 26901 \\
3 & 3 & 70 & 3 & 0.719 & 26973 \\
3 & 3 & 80 & 3 & 0.681 & 26978 \\\hline
3 & 3 & 20 & 4 & 0.666 & 26980 \\
3 & 3 & 30 & 4 & 0.671 & 26932 \\
3 & 3 & 40 & 4 & 0.681 & 27221 \\
3 & 3 & 50 & 4 & 0.691 & 27216 \\
3 & 3 & 60 & 4 & 0.662 & 27089 \\
3 & 3 & 70 & 4 & 0.689 & 27020 \\
3 & 3 & 80 & 4 & 0.699 & 27073 \\\hline
\end{tabular}
}
\end{center}

\begin{center}
{\tiny
\begin{tabular}{|c|c|c|c|c|c|}\hline
mod1 & mod2 & $\%$ Training set & Comp & AUROC & Time (ms)\\
 &  &  &  &  &  \\\hline
1 & 2 & 20 & 1 & 0.57 & 59887 \\
1 & 2 & 30 & 1 & 0.543 & 54270 \\
1 & 2 & 40 & 1 & 0.532 & 51231 \\
1 & 2 & 50 & 1 & 0.502 & 51748 \\
1 & 2 & 60 & 1 & 0.553 & 48869 \\
1 & 2 & 70 & 1 & 0.581 & 48208 \\
1 & 2 & 80 & 1 & 0.505 & 44480 \\\hline
1 & 2 & 20 & 2 & 0.913 & 60823 \\
1 & 2 & 30 & 2 & 0.894 & 58445 \\
1 & 2 & 40 & 2 & 0.915 & 53038 \\
1 & 2 & 50 & 2 & 0.925 & 54272 \\
1 & 2 & 60 & 2 & 0.883 & 49449 \\
1 & 2 & 70 & 2 & 0.892 & 53510 \\
1 & 2 & 80 & 2 & 0.905 & 45854 \\\hline
1 & 2 & 20 & 3 & 0.925 & 55210 \\
1 & 2 & 30 & 3 & 0.934 & 50355 \\
1 & 2 & 40 & 3 & 0.925 & 46006 \\
1 & 2 & 50 & 3 & 0.938 & 48730 \\
1 & 2 & 60 & 3 & 0.939 & 47193 \\
1 & 2 & 70 & 3 & 0.927 & 47459 \\
1 & 2 & 80 & 3 & 0.948 & 43050 \\\hline
1 & 2 & 20 & 4 & 0.777 & 56007 \\
1 & 2 & 30 & 4 & 0.768 & 54425 \\
1 & 2 & 40 & 4 & 0.777 & 49342 \\
1 & 2 & 50 & 4 & 0.801 & 50165 \\
1 & 2 & 60 & 4 & 0.73 & 50183 \\
1 & 2 & 70 & 4 & 0.773 & 55897 \\
1 & 2 & 80 & 4 & 0.764 & 49472 \\\hline
\end{tabular}
}
\end{center}

\begin{center}
{\tiny
\begin{tabular}{|c|c|c|c|c|c|}\hline
mod1 & mod2 & $\%$ Training set & Comp & AUROC & Time (ms)\\
 &  &  &  &  &  \\\hline
2 & 1 & 20 & 1 & 0.57 & 28224 \\
2 & 1 & 30 & 1 & 0.543 & 27507 \\
2 & 1 & 40 & 1 & 0.532 & 27992 \\
2 & 1 & 50 & 1 & 0.502 & 28657 \\
2 & 1 & 60 & 1 & 0.553 & 28987 \\
2 & 1 & 70 & 1 & 0.581 & 29142 \\
2 & 1 & 80 & 1 & 0.505 & 29956 \\\hline
2 & 1 & 20 & 2 & 0.908 & 27307 \\
2 & 1 & 30 & 2 & 0.89 & 27655 \\
2 & 1 & 40 & 2 & 0.907 & 27892 \\
2 & 1 & 50 & 2 & 0.917 & 27811 \\
2 & 1 & 60 & 2 & 0.88 & 29343 \\
2 & 1 & 70 & 2 & 0.888 & 28694\\
2 & 1 & 80 & 2 & 0.902 & 30529\\\hline
2 & 1 & 20 & 3 & 0.923 & 27011\\
2 & 1 & 30 & 3 & 0.936 & 27843\\
2 & 1 & 40 & 3 & 0.924 & 27659\\
2 & 1 & 50 & 3 & 0.934 & 27711\\
2 & 1 & 60 & 3 & 0.935 & 27715\\
2 & 1 & 70 & 3 & 0.912 & 27451 \\
2 & 1 & 80 & 3 & 0.947 & 27945\\\hline
2 & 1 & 20 & 4 & 0.777 & 27285\\
2 & 1 & 30 & 4 & 0.768 & 27341\\
2 & 1 & 40 & 4 & 0.777 & 27964\\
2 & 1 & 50 & 4 & 0.801 & 28070\\
2 & 1 & 60 & 4 & 0.73 & 28135\\
2 & 1 & 70 & 4 & 0.773 & 27540\\
2 & 1 & 80 & 4 & 0.764 & 28147\\\hline
\end{tabular}
}
\end{center}

\begin{center}
{\tiny
\begin{tabular}{|c|c|c|c|c|c|}\hline
mod1 & mod2 & $\%$ Training set & Comp & AUROC & Time (ms)\\
 &  &  &  &  &  \\\hline
3 & 1 & 20 & 1 & 0.534 & 27986   \\
3 & 1 & 30 & 1 & 0.556 & 27479   \\
3 & 1 & 40 & 1 & 0.547 & 28009  \\
3 & 1 & 50 & 1 & 0.534 & 29044  \\
3 & 1 & 60 & 1 & 0.559 & 31017  \\
3 & 1 & 70 & 1 & 0.595 & 29976  \\
3 & 1 & 80 & 1 & 0.562 & 33081  \\\hline
3 & 1 & 20 & 2 & 0.769 & 27458  \\
3 & 1 & 30 & 2 & 0.703 & 28111  \\
3 & 1 & 40 & 2 & 0.783 & 28059  \\
3 & 1 & 50 & 2 & 0.757 & 27854  \\
3 & 1 & 60 & 2 & 0.792 & 29617  \\
3 & 1 & 70 & 2 & 0.845 & 28810  \\
3 & 1 & 80 & 2 & 0.845 & 31435  \\\hline
3 & 1 & 20 & 3 & 0.785 & 27032  \\
3 & 1 & 30 & 3 & 0.81 & 27178   \\
3 & 1 & 40 & 3 & 0.831 & 27742  \\
3 & 1 & 50 & 3 & 0.876 & 27645  \\
3 & 1 & 60 & 3 & 0.779 & 27737  \\
3 & 1 & 70 & 3 & 0.831 & 27478  \\
3 & 1 & 80 & 3 & 0.84 & 28035   \\\hline
3 & 1 & 20 & 4 & 0.69 & 27143   \\
3 & 1 & 30 & 4 & 0.735 & 27257  \\
3 & 1 & 40 & 4 & 0.756 & 28155  \\
3 & 1 & 50 & 4 & 0.787 & 28089  \\
3 & 1 & 60 & 4 & 0.72 & 28012   \\
3 & 1 & 70 & 4 & 0.771 & 27695  \\
3 & 1 & 80 & 4 & 0.764 & 28079  \\\hline
\end{tabular}
}
\end{center}

\begin{center}
{\tiny
\begin{tabular}{|c|c|c|c|c|c|}\hline
mod1 & mod2 & $\%$ Training set & Comp & AUROC & Time (ms)\\
 &  &  &  &  &  \\\hline 
2 & 3 & 20 & 1 & 0.572 & 26908    \\
2 & 3 & 30 & 1 & 0.546 & 27512    \\
2 & 3 & 40 & 1 & 0.541 & 27206  \\
2 & 3 & 50 & 1 & 0.508 & 27139 \\
2 & 3 & 60 & 1 & 0.557 & 27412 \\
2 & 3 & 70 & 1 & 0.591 & 26957 \\
2 & 3 & 80 & 1 & 0.511 & 27141 \\\hline
2 & 3 & 20 & 2 & 0.909 & 26908 \\
2 & 3 & 30 & 2 & 0.891 & 27089 \\
2 & 3 & 40 & 2 & 0.904 & 27173 \\
2 & 3 & 50 & 2 & 0.915 & 27125 \\
2 & 3 & 60 & 2 & 0.88 & 27065  \\
2 & 3 & 70 & 2 & 0.888 & 26989 \\
2 & 3 & 80 & 2 & 0.901 & 26925 \\\hline
2 & 3 & 20 & 3 & 0.922 & 26984 \\
2 & 3 & 30 & 3 & 0.936 & 27064 \\
2 & 3 & 40 & 3 & 0.922 & 27153 \\
2 & 3 & 50 & 3 & 0.935 & 27287 \\
2 & 3 & 60 & 3 & 0.935 & 26959 \\
2 & 3 & 70 & 3 & 0.91 & 26954  \\
2 & 3 & 80 & 3 & 0.947 & 26973 \\\hline
2 & 3 & 20 & 4 & 0.776 & 26989 \\
2 & 3 & 30 & 4 & 0.768 & 26972 \\
2 & 3 & 40 & 4 & 0.779 & 27349 \\
2 & 3 & 50 & 4 & 0.803 & 27583 \\
2 & 3 & 60 & 4 & 0.73 & 27058  \\
2 & 3 & 70 & 4 & 0.774 & 26963 \\
2 & 3 & 80 & 4 & 0.768 & 27534 \\\hline
\end{tabular}
}
\end{center}

\begin{center}
{\tiny
\begin{tabular}{|c|c|c|c|c|c|}\hline
mod1 & mod2 & $\%$ Training set & Comp & AUROC & Time (ms)\\
 &  &  &  &  &  \\\hline 
1 & 3 & 20 & 1 & 0.557 & 11739 \\
1 & 3 & 30 & 1 & 0.564 & 9714 \\
1 & 3 & 40 & 1 & 0.534 & 9629 \\
1 & 3 & 50 & 1 & 0.531 & 9693 \\
1 & 3 & 60 & 1 & 0.572 & 9454 \\
1 & 3 & 70 & 1 & 0.579 & 9977 \\
1 & 3 & 80 & 1 & 0.527 & 9697 \\\hline
1 & 3 & 20 & 2 & 0.898 & 9403 \\
1 & 3 & 30 & 2 & 0.893 & 9531 \\
1 & 3 & 40 & 2 & 0.898 & 9573 \\
1 & 3 & 50 & 2 & 0.917 & 9387 \\
1 & 3 & 60 & 2 & 0.88 & 9617  \\
1 & 3 & 70 & 2 & 0.896 & 9388 \\
1 & 3 & 80 & 2 & 0.913 & 9331 \\\hline
1 & 3 & 20 & 3 & 0.927 & 9258 \\
1 & 3 & 30 & 3 & 0.94 & 9421  \\
1 & 3 & 40 & 3 & 0.925 & 9518 \\
1 & 3 & 50 & 3 & 0.939 & 9660 \\
1 & 3 & 60 & 3 & 0.941 & 9371 \\
1 & 3 & 70 & 3 & 0.928 & 9401 \\
1 & 3 & 80 & 3 & 0.943 & 9326 \\\hline
1 & 3 & 20 & 4 & 0.802 & 9219 \\
1 & 3 & 30 & 4 & 0.782 & 9300 \\
1 & 3 & 40 & 4 & 0.793 & 9681 \\
1 & 3 & 50 & 4 & 0.805 & 9632 \\
1 & 3 & 60 & 4 & 0.758 & 9577 \\
1 & 3 & 70 & 4 & 0.799 & 9399 \\
1 & 3 & 80 & 4 & 0.78 & 9527  \\\hline
 \end{tabular}
}
\end{center}

\begin{center}
{\tiny
\begin{tabular}{|c|c|c|c|c|c|}\hline
mod1 & mod2 & $\%$ Training set & Comp & AUROC & Time (ms)\\
 &  &  &  &  &  \\\hline 
3 & 2 & 20 & 1 & 0.537 & 13838 \\
3 & 2 & 30 & 1 & 0.572 & 12276 \\
3 & 2 & 40 & 1 & 0.556 & 12266 \\
3 & 2 & 50 & 1 & 0.545 & 12421 \\
3 & 2 & 60 & 1 & 0.557 & 11861 \\
3 & 2 & 70 & 1 & 0.597 & 11720 \\
3 & 2 & 80 & 1 & 0.564 & 10992 \\\hline
3 & 2 & 20 & 2 & 0.871 & 13036 \\
3 & 2 & 30 & 2 & 0.818 & 12823 \\
3 & 2 & 40 & 2 & 0.858 & 13585 \\
3 & 2 & 50 & 2 & 0.897 & 13735 \\
3 & 2 & 60 & 2 & 0.84 & 13357  \\
3 & 2 & 70 & 2 & 0.872 & 14152 \\
3 & 2 & 80 & 2 & 0.884 & 11571 \\\hline
3 & 2 & 20 & 3 & 0.886 & 12856 \\
3 & 2 & 30 & 3 & 0.865 & 13972 \\
3 & 2 & 40 & 3 & 0.898 & 14313 \\
3 & 2 & 50 & 3 & 0.932 & 15202 \\
3 & 2 & 60 & 3 & 0.9 & 14936   \\
3 & 2 & 70 & 3 & 0.919 & 15547 \\
3 & 2 & 80 & 3 & 0.922 & 13825 \\\hline
3 & 2 & 20 & 4 & 0.734 & 14685 \\
3 & 2 & 30 & 4 & 0.744 & 13716 \\
3 & 2 & 40 & 4 & 0.766 & 13411 \\
3 & 2 & 50 & 4 & 0.791 & 12101 \\
3 & 2 & 60 & 4 & 0.72 & 12054  \\
3 & 2 & 70 & 4 & 0.76 & 12297  \\
3 & 2 & 80 & 4 & 0.774 & 11715 \\\hline
 \end{tabular}
}
\end{center}

\end{document}